\documentclass[a4paper,fleqn,usenatbib]{mnras}

\usepackage{newtxtext,newtxmath}

\usepackage[T1]{fontenc}
\usepackage{ae,aecompl}

\usepackage{graphicx}	
\usepackage{amsmath}	
\usepackage{mathtools}	
\usepackage{array}	         
\usepackage{ragged2e}    
\usepackage{sidecap}                                         
\sidecaptionvpos{figure}{c}                                  
\usepackage[font=small, labelfont=bf]{caption}   
\usepackage{multirow}        				 
\usepackage{floatrow}                                         
\usepackage{arydshln}					 
\usepackage{bigstrut}					 
\usepackage{float}						
\usepackage[export]{adjustbox}                          

\graphicspath{ {./Figures/} }               
\numberwithin{equation}{section}    

\newcommand{\Msun}{{\,M}$_\odot$}

\newcommand{\Dsun}{{\,D}$_\odot$}

\newcommand{\kms}{{\,\rm {km s$^{-1}$}}}

\usepackage{longtable}  			
\usepackage{natbib}                          
\usepackage[autostyle]{csquotes}

\title[One Stream or Two - M31's NW Stream]{One Stream or Two - Exploring Andromeda's North West Stream}

\author[J. Preston et al]   
{Janet Preston,$^{1*}$
\ Denis Erkal,$^1$
\ Michelle L.M. Collins,$^1$
R. Michael Rich$^2$
\ Rodrigo Ibata,$^3$ 
Maxime Delorme$^4$
\\
\\
$^1$Department of Physics, University of Surrey, Guildford, GU2 7XH, Surrey, UK. \thanks{j.preston@surrey.ac.uk} \\
$^2$Department of Physics and Astronomy, UCLA, 430 Portola Plaza, Box 951547, Los Angeles, CA 90095-1547, USA \\
$^3$Observatoire de Strasbourg, 11, rue de l'Universit\'{e}, F-67000, Strasbourg \\
$^4$D\'{e}partement d’Electronique, des D\'{e}tecteurs et d’Informatique pour la Physique, CEA/IRFU, 91191 Gif-sur-Yvette, France  \\\\
}

\date{Accepted XXX. Received YYY; in original form ZZZ}

\pubyear{2019}

\begin{document}
\label{firstpage}
\pagerange{\pageref{firstpage}--\pageref{lastpage}}
\maketitle

\begin{abstract}
We present results of our dynamical stream modelling for the North West Stream in the outer halo of the Andromeda galaxy (M31).  Comprising two main segments, the North West Stream was thought to be a single structured arching around M31. However, recent evidence suggests that it is two separate, unrelated, streams. To test this hypothesis we use observational data from 6 fields associated with the upper segment of the North West Stream together with 8 fields and 5 globular clusters associated with the lower segment to constrain model orbits. We fit both segments of the stream using a fixed potential model for M31 and an orbit integrator to compare orbits with the observed streams. We measure the central tracks and predict proper motions for for the upper segment (lower segment) finding ${\mu^*_{\alpha}}$ = 0.078$^{+0.015}_{-0.012}$ (0.085$^{+0.001}_{-0.002}$) mas/yr and ${\mu_{\delta}}$ = $-$0.05$^{+0.008}_{-0.009}$ ($-$0.095$^{+0.003}_{-0.005}$) mas/yr. Our results support the hypothesis that the dwarf spheroidal galaxy Andromeda XXVII is the progenitor of the upper segment of the North West Stream and that the upper and lower segments do not comprise a single structure. We propose that the upper segment, which appears to be on an infall trajectory with M31, be renamed the \enquote{Andromeda XXVII Stream} and the lower segment, also apparently infalling towards M31, retain the name \enquote{North West Stream}.
\end{abstract}


\section{Introduction} \label{OM_Introduction}
\graphicspath{ {Figures/} }    

The sinuous spiral arms of the Andromeda Galaxy (M31) are surrounded by $\sim$30 satellite galaxies, more than 400 globular clusters (\citealt{RefWorks:627}) and $\sim$10 stellar streams and shell structures (\citealt{RefWorks:47}, \citealt{RefWorks:82}, \citealt{RefWorks:107}, \citealt{RefWorks:449}, {\citealt{M:977}). Remnants of long dead stellar systems accreted by their hosts, these streams and shells form when stars are stripped from a satellite under the influence of tidal disruption by a much larger host galaxy.  The elongation of a stream arises from the different energies acquired by the escaping stars producing \enquote{leading} and \enquote{trailing} tails about their disrupting progenitor, e.g. \cite{M:1047}, \cite{M:1080}, \cite{RefWorks:596}.   

The formation and perturbation of stellar streams has been well studied and has been the subject of work to: determine the gravitational potentials of galaxies (e.g. \citealt{RefWorks:645}, \citealt{RefWorks:176}, \citealt{RefWorks:10}, \citealt{RefWorks:421}, \citealt{M:1105}, \citealt{RefWorks:245}, \citealt{RefWorks:422}, \citealt{M:1015}, \citealt{M:1014}, \citealt{RefWorks:290}, \citealt{RefWorks:450}); explore galactic accretion history (e.g. \citealt{RefWorks:444, {RefWorks:38}}, \citealt{RefWorks:342}, \citealt{RefWorks:477}, \citealt{RefWorks:154},  \citealt{RefWorks:470}); determine the motion and properties of the stream progenitors (e.g.  \citealt{M:1114}, \citealt{RefWorks:163}, \citealt{RefWorks:646}, \citealt{M:924}) and search for the dark matter sub-haloes or missing satellites that have punched holes through or distorted the paths of streams over billions of years (e.g. \citealt{RefWorks:423}, \citealt{RefWorks:76}, \citealt{M:1046}, \citealt{RefWorks:327}, \citealt{M:1041, RefWorks:424, M:1013},  \citealt{M:1081}).

The North West (NW) Stream is an intriguing feature which probes a large radial extent of the M31 halo.  This stream of metal poor stars comprises two distinct segments, see K1 and K2 in Figure \ref{OM_Fig1}, with, currently, no visible spatial connection. Despite this, based on the morphology of the two segments at the time, \cite{RefWorks:18} described it as a single structure looped around M31. \cite{RefWorks:82} backed up this view by reporting similar photometric metallicities in the two segments. Earlier, \cite{RefWorks:76} found the stream to contain stars $\sim$10 Gyrs old, to be around $\sim$5 kpc wide and to cover a projected distance of $\sim$200 kpc, making it one of the longest streams in the Local Group. This analysis of the stream also noted that the lower segment (hereafter referred to as NW-K2) is virtually complete while the upper segment is not well-defined and has a number of obvious gaps, potentially consistent with the effects of dark matter sub-halos on the stream. 

\begin{figure*}
	\includegraphics[height=.55\paperheight, width=.8\paperwidth]{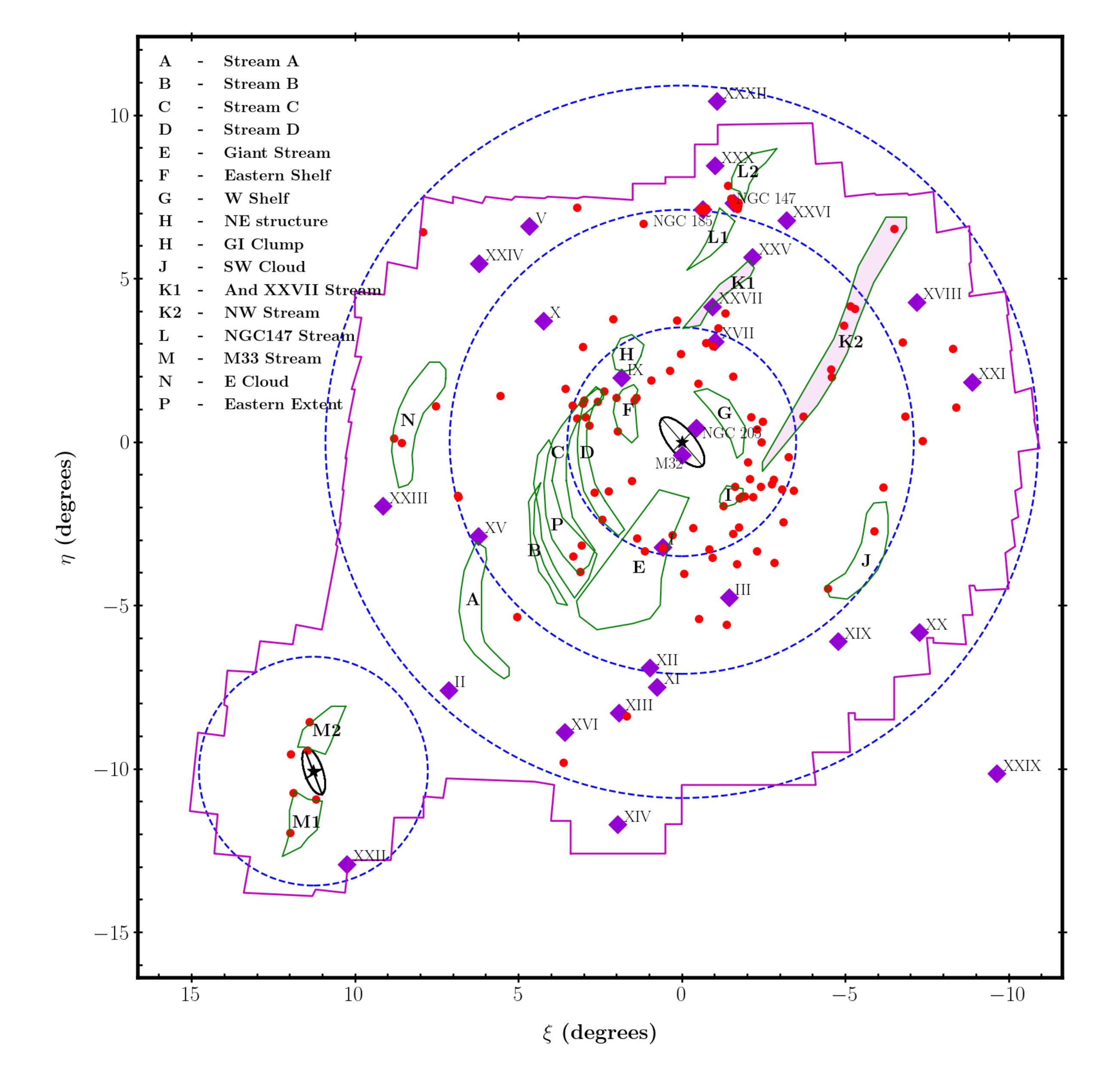}
    	\vspace*{-6mm}\caption[The PAndAS footprint]
	{The PAndAS footprint.  The purple icons represent M31 dwarf satellite galaxies (locations from \citealt{RefWorks:449}), the red icons denote the positions of globular clusters located at a projected radius $\sim$1$^\circ$ from the centre of M31 (locations from \citealt{RefWorks:516}). The blue dashed circles denote projected radii of 50kpc, 100kpc and 150kpc from M31, and 50kpc from M33.  The magenta polygon denotes the outline of the PAndAS footprint.
	The green polygons provide "by eye" indications of the locations of stellar streams and shells identified within the M31 halo.  They trace major features visible in Figure 11 of  \cite{RefWorks:449}.  Coordinates to plot these polygons were provided by Alan McConnachie (private communication) except for: stream B which was sourced from \cite{RefWorks:38}; the Eastern Extent, sourced from \cite{M:880} and stream K2, sourced from \cite{M:1147}.   
	}
	\label{OM_Fig1}
\end{figure*}

NW-K2 was discovered in the Pan-Andromeda Astronomical Survey (PAndAS,  \citealt{RefWorks:58}). It is $\sim$6$^{\circ}$ ($\sim$80 kpc) long in projection, with an estimated luminosity of \textit{M$_v$} = $-$12.3 $\pm $0.5,  \cite{RefWorks:449}, and is located $\sim$50 - 120 kpc from the centre of M31 extending away in a gentle curve, potentially to beyond the edge of the PAndAS footprint.  \cite{RefWorks:405} found NW-K2 to lie behind M31 with the northern part of the stream having a distance modulus of 24.63 $\pm$ 0.19 and the southern part lying some 20 kpc closer to us than the northern part. 

NW-K2 is also co-located on the sky with 7 globular clusters (GCs). The 6 furthest from M31 show a clear velocity trend indicative of a relationship between them (\citealt{RefWorks:95}, \citealt{RefWorks:72}, \citealt{RefWorks:232, RefWorks:236}). Work by \cite{RefWorks:76},  \cite{RefWorks:236}, \cite{RefWorks:516} and \cite{RefWorks:405} and \cite{M:1147} found strong kinematic evidence that many of the GCs are associated with the NW-K2 stream beyond the spatial co-location.  \cite{RefWorks:313} used this association of the GCs and the stream to model its orbit, determining a minimum perigalacticon of $\sim$25 kpc and estimating the half-light radius (\textit{r}$_{\rm h}$) of its progenitor to be $\ge$30pc. This implies its progenitor is too large to be a GC and more likely to be a dwarf galaxy with an estimated mass \mbox{$\ge$ 2.2 $\times$ 10$^6$ \Msun} which is consistent with \cite{RefWorks:449} who estimated a progenitor mass of $\sim$8.5 $\times$ 10$^6$ \Msun.  These estimates for a potentially large progenitor are backed up by \cite{RefWorks:516} whose work indicated that NW-K2 has a high specific frequency (which connects the total luminosity of a galaxy to the number of GCs it hosts), \textit{S$_N$} = 70 – 85, which could be explained by an additional association with the dwarf elliptical galaxies NGC 147 and NGC 185 lying to the north of the stream.

\begin{table*}
	\centering
	\setlength\extrarowheight{2pt}
	\caption[Properties for And XXVII and NW-K1 observed masks]
	{Properties for the observed fields across And XXVII and along NW-K1 and NW-K2, including: mask name; date observations were made; observing PI; Right Ascension and Declination of the centre of the mask, and the systemic velocities and number of confirmed members of the AndXXVII/NW-K1 and NW-K2 stellar populations for each mask. The $\alpha$ and $\delta$ for the centre of each mask are determined by taking the mean of the coordinates for all stars on the mask. The systemic velocities are the posterior values from MCMC analyses conducted for each mask, as described in P19 and  \cite{M:1147}, derived from stream stars only. The masks are listed in order of increasing distance from M31.
	}		
	\label{OM_table:1a}
	\begin{tabular}{lclcccccc} 
		\hline
		Mask name & Date & PI  & $\alpha_{\rm J2000}$ & $\delta_{\rm J2000}$  & v$_r$  &  Members \\ [0.5ex] 
		                  &          &       & $hh$ : $mm$ : $ss$     & $^o$ : $^{\prime}$ : $^{\prime \prime}$  & {\kms} &   \\ [0.5ex]
		\hline 
		NW-K1    &                      &                &                        &                           &                                                                &   \\ [1.5ex]   
		A27sf1    &  2015-09-12  &  Collins   &  00:39:39.96   & +45:08:47.73     &    $-$542.3  $^{+    7.1  }_{-    7.4  }$      &    8    \\ [1.5ex] 
		603HaS  &  2010-09-09  &  Rich       &  00:38:58.52   & +45:17:32.20     &    $-$530.2  $^{+    2.9  }_{-    3.1  }$      &    8   \\ [1.5ex]  
		7And27   &  2011-09-26  &  Rich       &  00:37:29.40   & +45:24:12.50     &    $-$526.1  $^{+   10.0  }_{-   11.0  }$    &   11  \\ [1.5ex]  
		A27sf2    &  2015-09-12  &  Collins   &  00:36:13.17   & +45:32:31.68     &    $-$518.4  $^{+   12.5  }_{-   12.5  }$    &   2  \\ [1.5ex]   
		604HaS  &  2010-09-09  &  Rich       &  00:32:05.16   & +46:08:31.20     &    $-$507.4  $^{+   10.7  }_{-   10.8  }$    &    1  \\ [1.5ex] 
		A27sf3    &  2015-09-12  &  Collins   &  00:30:25.60   & +46:14:52.66     &    $-$498.6  $^{+    5.3  }_{-    5.3  } $     &    4  \\ [1.5ex]  
		
		NW-K2    &                       &                &                        &                           &                                            &   \\ [1.5ex]  
		NWS6     &  2013-09-12   &  Mackey  &  00:28:25.83   &  +40:45:54.74   &   $-$435.8 $\pm$ 19.8        &        11   \\[1.5ex]  	
		NWS5     &  2013-09-11   &  Mackey  &  00:20:03.15   &  +42:44:18.98   &   $-$442.5 $\pm$ 19.1        &        6   \\[1.5ex] 		
		507HaS  &  2009-10-15   &  Rich       &  00:17:58.08   &  +43:07:0.14     &   $-$439.9 $\pm$ 24.1        &        3   \\[1.5ex] 
		506HaS  &  2009-10-15   &  Rich       &  00:15:58.09   &  +43:58:48.47   &   $-$431.5 $\pm$ 6.7          &        3   \\[1.5ex]  		
		NWS3     &  2013-09-11   &  Mackey  &  00:13:33.45   &  +44:43:08.81  &   $-$454.1 $\pm$ 11.9         &        8   \\[1.5ex]   		
		704HaS  &  2011-09-27   &  Rich       &  00:11:03.27   &  +45:32:44.0     &   $-$438.5 $\pm$ 8.3          &        4   \\[1.5ex]   
		606HaS  &  2010-09-09   &  Rich       &  00:08:36.35   &  +46:38:36.04   &  $-$438.4 $\pm$ 0.0          &        1   \\[1.5ex]  	
		NWS1    &  2013-09-12   &  Mackey  &  00:07:56.18   &  +47:06:42.68   &   $-$430.9 $\pm$ 9.1.        &       3   \\[1.5ex] 
		\hline
	\end{tabular}
\end{table*}

The upper segment, hereafter referred to as NW-K1, was discovered along with its plausible progenitor, Andromeda XXVII (And XXVII), by \cite{RefWorks:18}.  They found NW-K1 to be $\sim$3$^{\circ}$ ($\sim$40 kpc) long in projection, located $\sim$50 - 80 kpc from M31's centre. Later work by \cite{RefWorks:449} estimated the the stellar mass of the stream to be 9.4 × 10$^5$\Msun {  }with a luminosity of \mbox{\textit{M$_v$} = $-$10.5 $\pm $0.5}.  Kinematic analysis by \cite{RefWorks:42} led these authors to agree with \cite{RefWorks:18} that And XXVII is not in dynamical equilibrium and is no longer a bound system, a view also supported by  \cite{RefWorks:206}, \cite{RefWorks:474}  and \cite{RefWorks:586} (hereafter P19), who found And XXVII to be a plausible candidate for the progenitor of NW-K1. P19 also found evidence that the NW Stream may not be a single structure. They detected a velocity gradient of \mbox{$-$1.7$\pm$0.3 {\kms} kpc$^{-1}$} along NW-K1 which taken together with the systemic velocities of And XXVII and NW-K1 they considered to be indicative of an infall trajectory. Comparing this result with that of \cite{RefWorks:236}, who found a velocity gradient of \mbox{$-1.0$$\pm$0.1{\kms} kpc$^{-1}$} across the globular clusters on NW-K2, also potentially indicative of an infall trajectory towards M31, it seemed unlikely that the two segments were part of a single structure.

To test this hypothesis and that of And XXVII being a viable candidate for progenitor of NW-K1, we model both segments of the NW stream with the aim of:
\begin{itemize}
\item Simulating the stream produced by And XXVII by modelling it as an orbit.
\item Modelling orbits that match the track of the NW-K2 stream.  We will examine these orbits, along with those for NW-K1, to see if there are any that connect the two streams. 
\item Obtaining predictions for the proper motions for NW-K1 and NW-K2.
\item Determining pericentric distances for NW-K1 and NW-K2.
\end{itemize}

The paper is structured as follows: Sections \ref{OM_Observations} and \ref{OM_Modelling Approach} describe our observational data and approach to simulating and fitting the orbits of the progenitors of NW-K1 and NW-K2. Section \ref{OM_Discussion}  presents a discussion of our findings.  Our conclusions  are presented in Section \ref{OM_Conclusions}.

\begin{figure*}
  	\centering
	\includegraphics[height=.34\paperheight, width=.75\paperwidth]{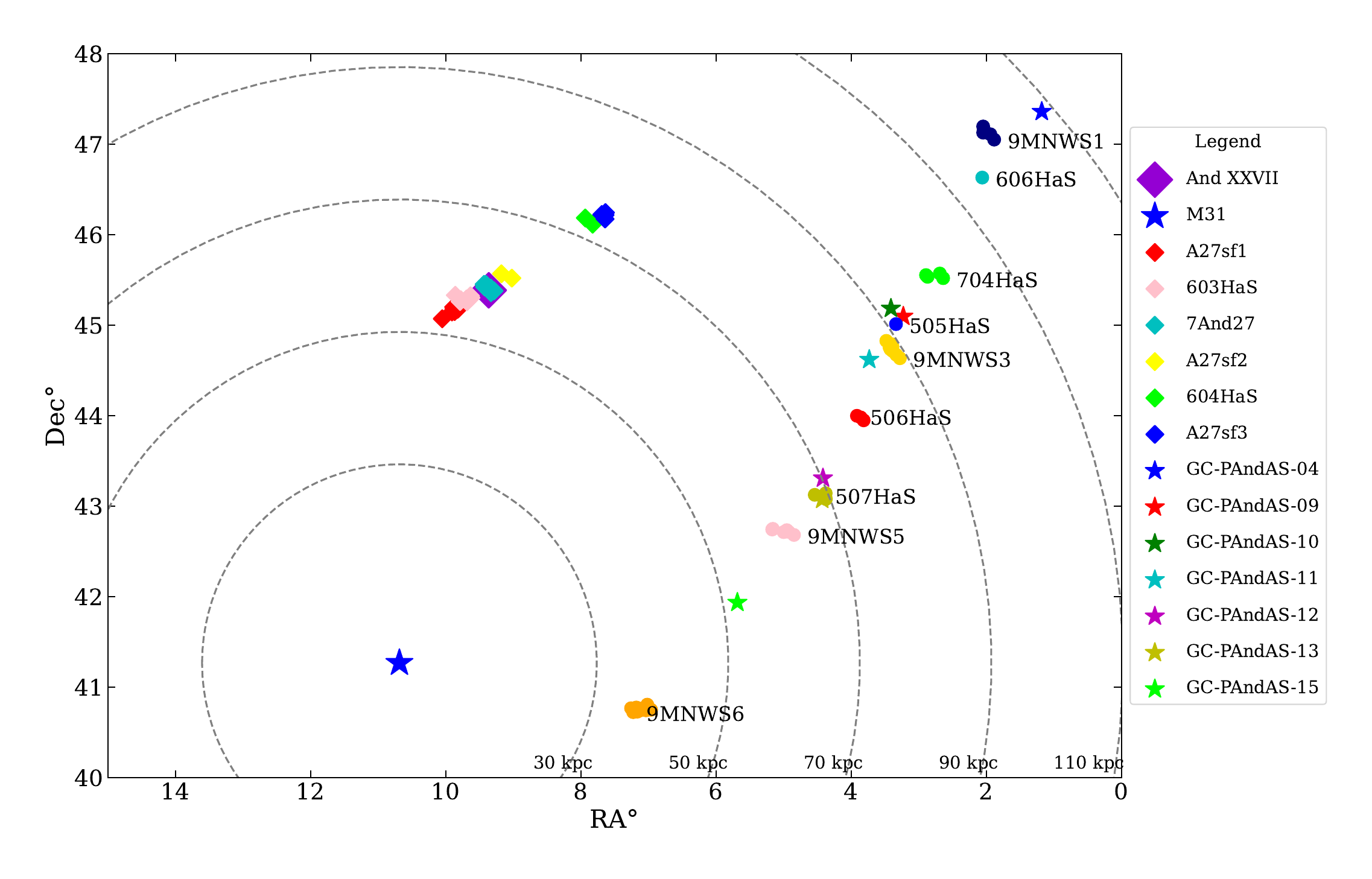}
    	\vspace*{-6mm}\caption[On-sky positions of And XXVII, NW-K1 and observed masks]
	{On-sky positions of And XXVII, GCs and observing masks for both NW-K1 and NW-K2.  The dotted lines show radial distances from the centre of M31 from 30 - 130 kpc (the outer most dotted line in the top right hand corner of the plot).
	}
	\label{OM_Fig2}
\end{figure*}

\begin{table}
	\centering
	\setlength\extrarowheight{2pt}
	\caption[Properties for NW-K2 globular clusters]
	{Properties for NW-K2 globular clusters from \cite{RefWorks:236}, \cite{RefWorks:516}.}		
	\label{OM_table:1}
	\begin{tabular}{lccc} 
		\hline
		Name & $\alpha_{\rm J2000}$ & $\delta_{\rm J2000}$ & v$_r$ \\ [0.5ex]
		           & $hh$ : $mm$ : $ss$     & $^o$ : $^{\prime}$ : $^{\prime \prime}$  & {\kms} \\ [0.5ex]
		\hline 
		PAndAS-04          & 00:04:42.90  & +47:21:42.00   &   $-$397 $\pm$ 7     \\ [0.5ex] 
		PAndAS-09          & 00:12:54.60  & +45:05:55.00   &.  $-$444 $\pm$ 21   \\[0.5ex] 
		PAndAS-10          & 00:13:38.60  & +45:11:11.00   &   $-$435 $\pm$10     \\[0.5ex] 
		PAndAS-11          & 00:14:55.60  & +44:37:16.00   &   $-$447 $\pm$ 13    \\[0.5ex] 
		PAndAS-12          & 00:17:40.00  & +43:18:39.00  &   $-$472 $\pm$ 5       \\[0.5ex] 
		PAndAS-13         &  00:17:42.7    & +43:04:31.0    &   $-$570 $\pm$ 45     \\[0.5ex]  
		PAndAS-15         &  00:22:44.0    & +41:56:14.0    &   $-$385 $\pm$ 6       \\[0.5ex]  
		\hline
	\end{tabular}
\end{table}

\section{Observations} \label{OM_Observations}
\graphicspath{ {Figures/} } 

The photometric data were obtained as part of the Pan-Andromeda Archaeological Survey (PAndAS, \citealt{RefWorks:58}).  This used the \mbox{3.6 m} Canada-France-Hawaii Telescope (CFHT) with the MegaPrime/MegaCam camera, comprising 36, 2048 x 4612, CCDs with a pixel scale of 0.185 arcsec/pixel able to deliver $\sim$1 degree$^2$ field of view (\citealt{RefWorks:58}). \textit{g}-band (4140\AA{}- 5600\AA) and \textit{i}-band (7020\AA{}-8530\AA) filters were used to ensure good colour discrimination of red giant branch (RGB) stars. With good seeing of < 0.8 arcsec, individual stars were resolved to depths of \textit{g} = 26.5 and \textit{i} = 25.5 with a signal to noise ratio $\sim$10 (\citealt{RefWorks:58}, \citealt{RefWorks:42}, \citealt{RefWorks:81}).   

The data were first reduced at CFHT using the Elixir system, \cite{RefWorks:570}.  This process ascertained the photometric zero points then de-biased, flat-fielded and fringe-corrected the data.   Next, the  Cambridge Astronomical Survey Unit processed the data using a bespoke pipeline described in \cite{RefWorks:336}. The data were then classified morphologically as, e.g., point source, non-point source and noise-like, and stored with the \textit{g} and \textit{i} data (see \citealt{RefWorks:18}). 

Follow-up observations were conducted to obtain spectroscopic data for 6 fields across the centre of And XXVII and along the length of NW-K1 and 8 fields along NW-K2, see Table \ref{OM_table:1a} and Figure \ref{OM_Fig2}.  These observations used the DEep-Imaging Multi-Object Spectrograph (DEIMOS) on the Keck II Telescope with the OG550 filter and 1200 lines/mm grating with a resolution of $\sim$1.1\AA {}-1.6\AA {} at FWHM. The fields were observed as follows: NWS3 and NWS6 were observed for a total of 1 hour 40 mins split into 5 x 20 minute integrations; 506HaS, 507HaS and 704HaS were observed for 1 hour 30 mins (3 x 30 minutes); NWS1 and NWS5 had a total observation time of 1 hour 20 minutes (4 x 20 minutes); all the NW-K1 fields were observed for 1 hour (3 x 20 minutes) and 606HaS had 3 x 15 minute integrations with a total observation time of 45 minutes.
 
We selected targets stars based on their location within the colour magnitude diagram (CMD). The highest priority were bright stars which lay directly on the And XXVII/NW-K1 and NW-K2 RGBs with 20.3 < \textit{i$_0$} < 22.5 (where \textit{i$_0$} is the de-reddened \textit{i}-band magnitude, given by \textit{i$_0$} = \textit{i} - 2.086E(B-V), obtained using extinction maps and correction coefficient, from Table 6, in \cite{M:1104}.  Next, we prioritised fainter stars on the RGBs, i.e. 22.5 < \textit{i$_0$} < 23.5. We filled the remainder of the mask with stars in the field where 20.5 < \textit{i$_0$} < 23.5 and 0.0 < \textit{g-i} < 4.0.

We reduced the data using a specifically constructed pipeline described in \cite{RefWorks:216}.  This pipeline corrected for: scattered light, flat-fields, the slit function and illumination within the telescope as well as calibrating the wavelength of each pixel. The pipeline also determined the velocities and associated uncertainties for the stars by: creating model spectra comprising a continuum and the absorption profiles of the Calcium Triplet (CaT) lines (at 8498\AA, 8542\AA{} and 8662\AA); cross-correlating these models with non-resampled stellar spectra to obtain the Doppler shift and CaT line widths; and correcting the velocities and associated uncertainties to the heliocentric frame.

P19 described the approach to confirming secure stellar populations for And XXVII and NW-K1 and \cite{M:1147} describe the same for NW-K2.  Both approaches included obtaining an overall probability of membership based on each star's radial velocity and its proximity to a fiducial isochrone overlaying the And XXVII/NW-K1 and NW-K2 RGBs on the CMD. The numbers of confirmed stars on each mask are shown in Table \ref{OM_table:1a} and their properties are included in Appendix \ref{Properties of stream stars}. In addition to the confirmed stellar population for NW-K2, we also use the properties of the GCs PAndAS-04, PAndAS-09, PandAS-10, PandAS-11 and PAndAS-12, co-located on-sky (\citealt{RefWorks:236}), see Table \ref{OM_table:1} and Figure \ref{OM_Fig2}. These are consistent with those used by \cite{RefWorks:313} and were subsequently determined by \cite{RefWorks:405} and \cite{M:1147} to have similar properties to those of NW-K2, which supports the concept that they originated from the same progenitor as the stream.

Throughout this work we adopt an heliocentric distance of \mbox{827 $\pm$ 47 kpc} (\citealt{RefWorks:18}) and a systemic velocity of $-$526.1$^{+10.0}_{-11.0}${\kms} (P19) for And XXVII. We also assume a radial velocity of $-$300 $\pm$ 4 {\kms} and an heliocentric distance of 783 $\pm$ 25 kpc and for M31 (\citealt{RefWorks:56}) .  With respect to this latter value we note that it precedes the current findings for M31's heliocentric distance of 761 $\pm$ 11 kpc by \cite{M:899} and 798 $\pm$ 28 kpc by \cite{M:1091} but recognise that, as it is consistent with and has been used in the determination of the properties and other values used within this paper, it is appropriate for use in our analysis.  We use values determined by  \cite{M:838}  for the proper motion of M31, i.e. ${\mu^*_{\alpha}}$ = 0.049 $\pm$ 0.010 mas/yr (where ${\mu^*_{\alpha}}$ = ${\mu_{\alpha}}$ cos$\delta$ mas/yr) and ${\mu_{\delta}}$ = $-$0.037 $\pm$ 0.008 mas/yr.

\section{Modelling Approach} \label{OM_Modelling Approach}
 
\subsection{Stream Models} \label{OM_Stream Models}

\subsubsection{Simulation Set-up} \label{OM_Sim_Set_up}
While tidal streams do not exactly follow an orbit (\citealt{RefWorks:641}), in many cases orbits can be used as simple models for streams. Given the modest amount of data that we have for NW-K1 and NW-K2 this motivates the use of simpler orbit models to trace the track of the stream as done by, for example, \cite{RefWorks:423, RefWorks:176}, \cite{RefWorks:421, RefWorks:537, M:1003}, \cite{M:1105}, \cite{RefWorks:422} and  \cite{RefWorks:468}, so we adopt this as a working assumption for our models. Given that P19 found And XXVII to be a plausible contender for the progenitor of NW-K1, if we model its orbit we should obtain an acceptable model of the NW-K1 stream track. 

To create the model orbits for the stream we convert the $\alpha$, $\delta$ and velocity for M31 and our stream progenitors (i.e. And XXVII for NW-K1 and PAndAS-12 for NW-K2, following the approach by \citealt{RefWorks:313}) to galactocentric coordinates which, along with the other observables (i.e. distance, ${\mu^*_{\alpha}}$ and ${\mu_{\delta}}$) we then convert into 3d positions and velocities. We determine the position and velocity vectors for the streams' progenitors relative to M31 by subtracting the M31 position and velocity vectors from those of the progenitors.  We then create simulations of the stream orbit using a leapfrog integrator to generate possible trajectories both backwards and forwards from the location of the progenitor.  

We model the potential of M31 using parameters reported by \cite{ RefWorks:155, RefWorks:245}, as their models for other structures around M31 have yielded results consistent with observed data.  As shown in Table \ref{OM_table:3}, we model the bulge as an Hernquist sphere (\citealt{RefWorks:685}) with a mass of 3.24 $\times$  10$^{10}${\Msun} and \textit{r}$_h$ = 0.61 kpc.  We select a Miyamoto-Nagai disk (\citealt{RefWorks:683}) with a mass of 7.34 $\times$ 10$^{10}${\Msun} and \textit{r}$_{\rm h}$ = 5.94 kpc. We treat the disk as spherical (hence the scale height, or b parameter, = 0 kpc) as we are only interested in orbits that lie far from the M31 disk and so that we do not need to rotate our disk model to align with M31's disk.  Finally we define an NFW halo (\citealt{RefWorks:684}) with a mass of 1.995 $\times$ 10$^{12}${\Msun}, \textit{r}$_h$ = 28.9 kpc, and a concentration of 8.9, \cite{ RefWorks:155},  to represent the M31 halo.  Once the model orbit is generated we convert its position and velocity data back to heliocentric values for use in the  $\chi$$^2$ analysis and subsequent modelling. 

\begin{table}
	\centering
	\setlength\extrarowheight{2pt}
	\caption[Parameters for modelling the M31 potential]
	{Parameters for modelling the M31 potential.  All values were obtained and/or derived from  \cite{RefWorks:155, RefWorks:245} 
	}		
	\label{OM_table:3}
	\begin{tabular}{ll} 
		\hline
		Parameter &  Value \\
		\hline 
Hernquist Bulge          &          \\
Bulge mass                 & 3.24 $\times$  10$^{10}${\Msun}  \\
Scale radius                & 0.61 kpc \\
\\ [-2ex] 
Miyamoto-Nagai disk   &          \\
Disk mass                    & 7.34 $\times$  10$^{10}${\Msun}  \\
Scale radius                & 5.94 kpc \\
Scale height                & 0 kpc    \\
\\ [-2ex]
NFW halo                   &          \\
Halo mass                  & 1.995 $\times$  10$^{12}${\Msun}  \\
Scale radius               & 28.9 kpc \\
Halo concentration     & 8.9   \\[1ex]  
		\hline
	\end{tabular}
\end{table}

In all the analyses for NW-K1 and NW-K2, our key assumptions for fitting the streams are:
\begin{itemize}
\item We can fix the potential of M31, as defined in Table \ref{OM_table:3}, and determine values for the other parameters, ${\mu^*_{\alpha}}$,  $\mu$$_{\delta}$, radial velocity and distance, to produce orbits that follow the observed track of the stream.
\item The stream follows the orbit.  This is reasonable given the quality of the data and for stream progenitors likely to be low mass dwarf galaxies. 
\item We can ignore the effect of dynamical friction on the orbit.  We do this for (a) simplicity, (b) because the mass of And XXVII is unknown and (c) because it is likely to have only a small  impact on the energy of the stream, \cite{RefWorks:351}.
\item There is no interaction from any of the other M31 satellites. 
\end{itemize}

\subsubsection{Modelling NW-K1} \label{OM_Modelling_NW-K1}

As we do not have the six-dimensional data required to generate accurate orbits for the stream tracks, we begin our modelling with a $\chi$$^2$ analysis on a grid of orbits to determine initial values for proper motions for And XXVII. Our models of the stream assume that the generated orbits will trace the centre-line of the NW-K1 stream track. We initiate the orbits using the properties of  And XXVII \mbox{($\alpha$ = 00$^h$ 37$^m$ 27$^s$.1} and $\delta$ =  +45$^\circ$: 23$^m$ 13$^s$.0) as the starting point.  We randomly sample the velocity parameter from a gaussian with a mean and a dispersion.  We define the initial value of the mean to be the systemic velocity of And XXVII, v$_r$ = $-$526.1$^{+10.0}_{-11.0}${\kms} (P19) and obtain the initial value of the dispersion by combining in quadrature the uncertainties for the systemic velocity of And XXVII and M31 (as we have fixed its radial velocity).  We initialise the distance parameter to the heliocentric distance for And XXVII taking random samples within a dispersion defined by combining in quadrature the relevant uncertainties for And XXVII and M31.

Given that proper motions are not available for And XXVII, we sample random values from a normal distribution around ${\mu^*_{\alpha}}$ = 0.0 $\pm$ 0.059 mas/yr and ${\mu_{\delta}}$ = 0.0 $\pm$ 0.059 mas/yr, which equates to a velocity dispersion at the distance of M31 of $\sim$218 {\kms} that is consistent with the relative velocity of And XXVII/NW-K1 and M31.  We convert the velocity dispersion using:
\begin{equation} 
	\label{OM_eq:001a}
		\mu  = v_{\rm{rel}} / (4.74 ${\textit{\Dsun}}$_{\rm{M31}})
\end{equation}
where $\mu$ represents ${\mu^*_{\alpha}}$ or ${\mu_{\delta}}$ mas/yr, \textit{v}$_{\rm{rel}}$ \kms{ }is the velocity dispersion of M31, 4.74 is a unit conversion factor, and {\textit{\Dsun}}$_{\rm{M31}}$ is the heliocentric distance of M31 in kpc.  

We use the above parameters to generate 50,000 stream models each of which is integrated forwards and backwards for 0.5 Gyrs in time steps of 0.1 Myrs, which is sufficient time to generate orbits that are significantly longer than the observed streams we will compare them with and should provide a sufficient timeframe to detect any credible connections between the two streams.

\begin{figure*}
 	\centering
	\includegraphics[ width=.86\paperwidth]{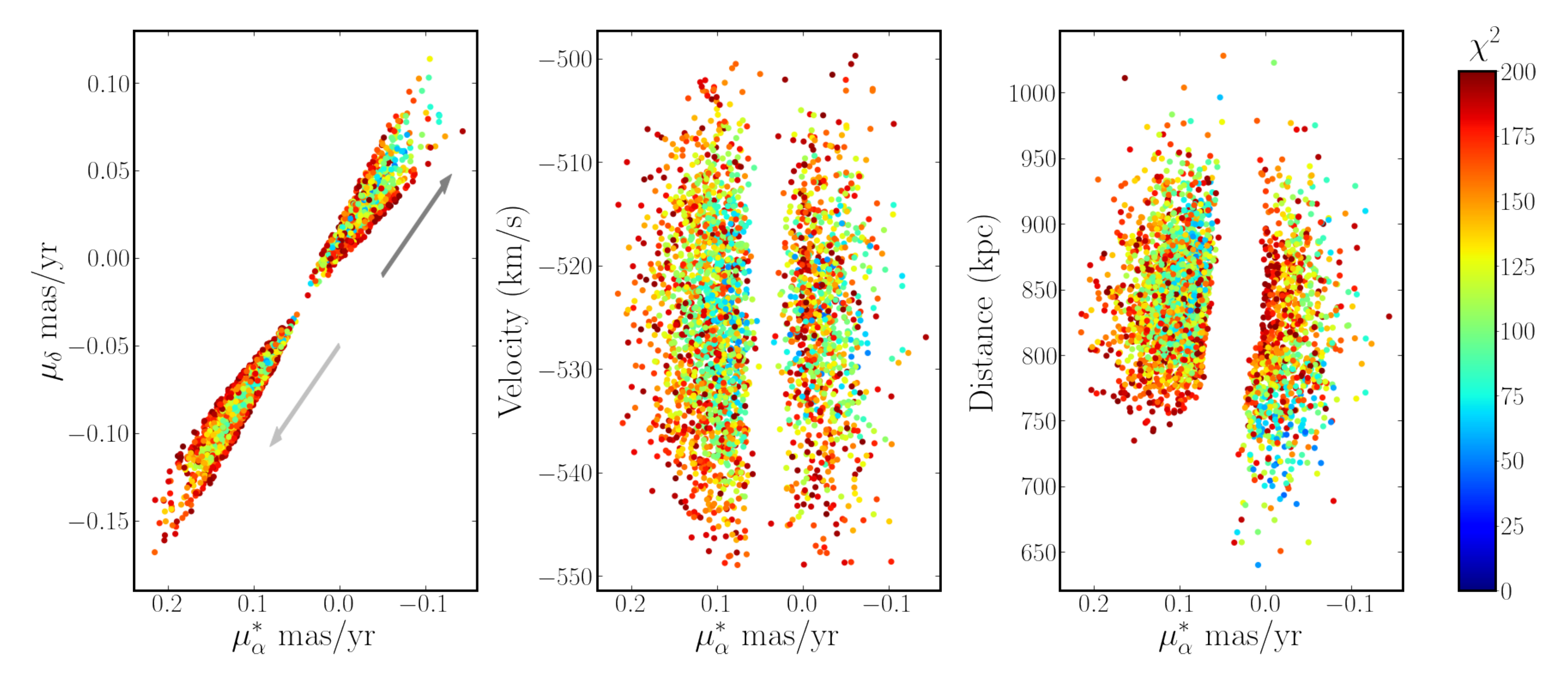}
    	\vspace*{-7mm}\caption[$\chi$$^2$  analysis of the stream models for NW-K1 ]
	{$\chi$$^2$ analysis of the stream models for NW-K1 showing only the high likelihood solutions with $\chi$$^2$ < 200 to determine potential proper motions for And XXVII.  The left hand panel, ${\mu_{\delta}}$ versus ${\mu^*_{\alpha}}$, indicates that there are two possible solutions for these values, which would indicate that there are two possible orbits that could fit the stream track i.e. one moving away from (indicated by the dark grey arrow) and one moving towards M31 (indicated by the light grey arrow). The middle panel, And XXVII radial velocity versus ${\mu^*_{\alpha}}$, shows the radial velocities sampled during the analysis. The right hand plot, And XXVII heliocentric distance versus ${\mu^*_{\alpha}}$, also shows the distances sampled during the analysis. 
	}
	\label{OM_Fig_3}
\end{figure*}  

\begin{figure*}
 	\centering
	\includegraphics[ width=.86\paperwidth]{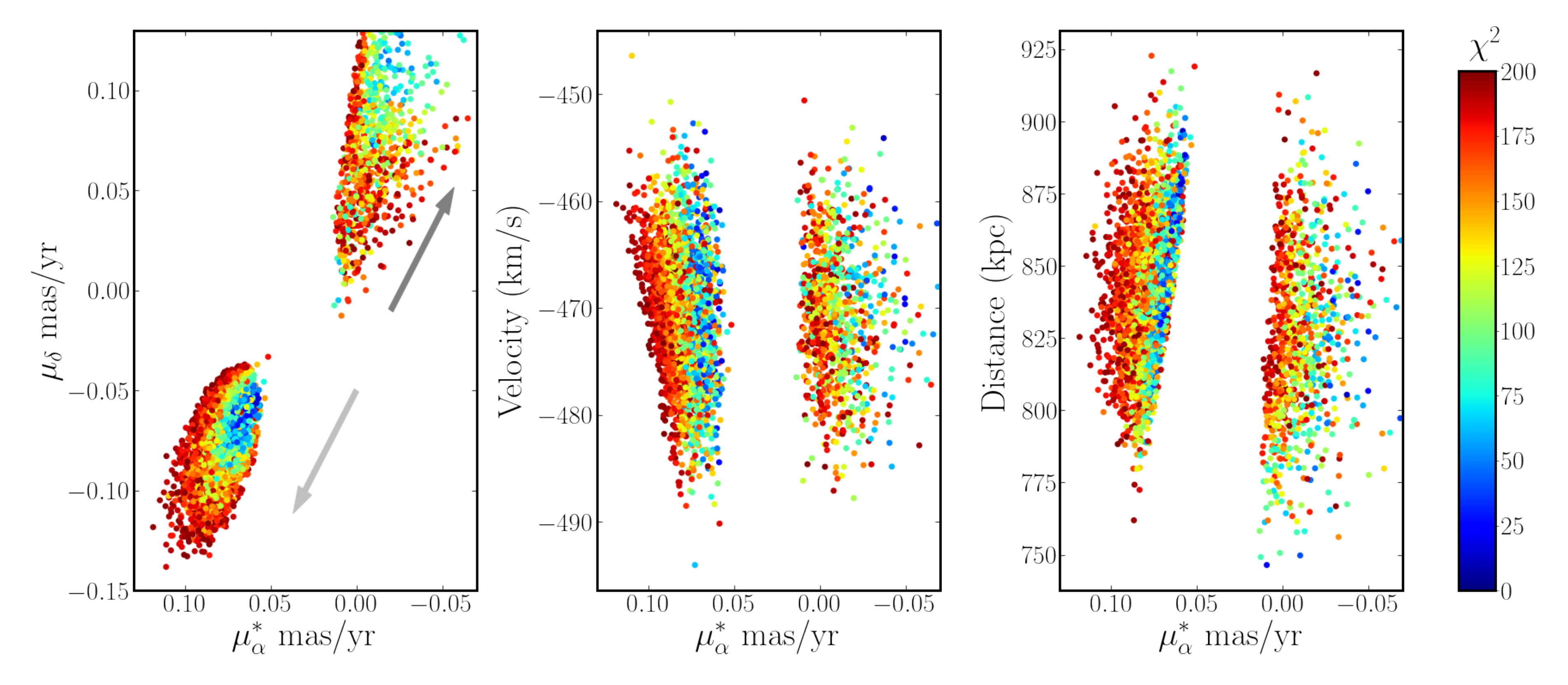}
    	\vspace*{-7mm}\caption[$\chi$$^2$ analysis of the stream models for NW-K2 ]
	{$\chi$$^2$ analysis of the stream models for NW-K2 showing only the high likelihood solutions with $\chi$$^2$ < 200 to determine potential proper motions for PAndAS-12. The middle and right hand panels show the systemic velocity of PAndAS-12 as a function of ${\mu^*_{\alpha}}$ and the heliocentric distance of PAndAS-12 as a function of ${\mu^*_{\alpha}}$ respectively.  The left hand panel shows ${\mu_{\delta}}$ versus ${\mu^*_{\alpha}}$ and indicates that there are two possible solutions for the proper motions for NW-K2  i.e. one for the stream moving away from (indicated by the dark grey arrow) and one with the stream moving towards M31 (indicated by the light grey arrow).  With the highest density of low $\chi^2$ values appearing in this latter region of the plot, this could be indicative of a preference for the motion of NW-K2 being towards M31, which is consistent with findings by \cite{RefWorks:313}.
	}
	\label{OM_Fig_10}
\end{figure*}

To find the best fit orbit we compare each stream model to the data.  In particular we conduct a $\chi$$^2$  analysis for the declination and line of sight velocity using the centres of the masks shown in Table \ref{OM_table:1a}. To find the orbit with properties that most closely resemble our mask data we first find the two right ascensions in the orbit $\alpha$$^o_l$ and $\alpha$$^o_r$ that lie either side of the closest match to the centre of each mask ($\alpha_{f, \textit{j}}$).  We then obtain the corresponding values for the declination ($\delta$$^o_l$ and $\delta$$^o_r$) and the radial velocity (v$^o_{r,l}$ and v$^o_{r,r}$) and use linear interpolation, in $\alpha$, to obtain the values from the model orbit that are the closest match to the observed data. We then calculate the respective $\chi$$^2$ for each of the properties using:

\begin{equation} 
	\label{OM_eq:004}
		\chi^2_{\delta}  = \sum_{i}\bigg(\frac{\delta_{\rm o, i} - \delta_{\rm f, j} } {\sigma_{\delta, \rm f} } \bigg)^2 ,
\end{equation}

\begin{equation} 
	\label{OM_eq:005}
		\chi^2_{v}  = \sum_{i}\bigg(\frac{v_{\rm o, i} - v_{\rm f, j} } {\sigma_{\rm v, f} } \bigg)^2 ,
\end{equation}

\begin{equation} 
	\label{OM_eq:006}
		\chi^2_{d}  = \sum_{i}\bigg(\frac{d_{\rm o, i} - d_{\rm f, j} } {\sigma_{\rm A27} } \bigg)^2 ,
\end{equation}

\noindent where: $\chi^2_{\delta}$, $\chi^2_v$ and $\chi^2_d$ are the $\chi^2$ values for the positional, velocity and distance parameters; $\sigma_{\delta, \rm f}$} is the uncertainty on the positional parameter, for which we use the standard deviation from the mean of the $\alpha$ and $\delta$ values for the centre of the mask, $\sigma_{\rm v, f}$ is the uncertainty on the systemic velocity value calculated by P19 and $\sigma_{\rm A27}$ is the distance uncertainty, obtained from \cite{RefWorks:18}. Finally, we combine the above $\chi$$^2$ values to find an overall total for the model that best matched all the observed properties collectively i.e.
\begin{equation} 
	\label{OM_eq:006a}
		\chi^2_{\rm tot}  = \chi^2_{\delta}  + \chi^2_{v}
\end{equation}

\noindent As there are no reliable distances to the centres of the masks $\chi^2_d$ was not included in the final determination of the overall $\chi^2$. 

Table \ref{OM_table:30} and Figure \ref{OM_Fig_3} show the results of our $\chi$$^2$ analysis, which indicates that there are two possible solutions for the 2-dimensional proper motion of NW-K1. The scatter at the top right of the left hand panel of Figure \ref{OM_Fig_3} is consistent with motion in direction from south-east to north-west (in the same frame of reference as in Figure  \ref{OM_Fig1}). For simplicity, throughout the rest of this paper we refer to this motion as \enquote{away from} M31, for both NW-K1 and NW-K2.  
The scatter at the bottom left of this plot is consistent with motion in a direction from north-west to south east,  Again, for the purposes of simplicity we will, henceforth, refer to motion in this direction as being \enquote{to} or \enquote{towards} M31. In Section \ref{OM_Orientation} we will confirm that towards and away do actually mean radially towards and radially away from M31.

\begin{table}
	\centering
	\setlength\extrarowheight{2pt}
	\caption[Lowest $\chi$$^2$ values]
	{$\chi$$^2$ values for the best fit orbits of NW-K1 and NW-K2 moving in both directions. 
	}		
	\label{OM_table:30}
	\begin{tabular}{lc} 
		\hline
		Stream &  $\chi$$^2$ \\
		\hline 
NW-K1 towards M31     & 45.69 \\
NW-K1 away from M31 & 48.32 \\
NW-K2 towards M31     & 11.96 \\
NW-K2 away from M31 & 12.09 \\			
\hline
	\end{tabular}
\end{table}

\subsubsection{Modelling NW-K2} \label{OM_Modelling_NW-K2}
To produce model orbits for NW-K2, we follow the process described in Section \ref{OM_Sim_Set_up}, adapted in line with approaches described by \cite{RefWorks:313} and \cite{RefWorks:405}.  We assign the properties, $\alpha$ = 00$^h$ 17$^m$ 40$^s$.0,  $\delta$ = 43$^\circ$ 18$^m$ 39$^s$ and line of sight radial velocity, \textit{v} = $-$472 $\pm$ 5 {\kms} of globular cluster PAndAS-12 (\citealt{RefWorks:236}) which we assume, as did \cite{RefWorks:313}, to be the centre of the stream.   For the purposes of our models we define the heliocentric distance for our stream centre to be ${\textit{\Dsun}}$ $\approx$ 834 $\pm$ 9.0 kpc, this being the average of distances to four locations along NW-K2 determined by \cite{RefWorks:405}.

We run the simulator to generate 50,000 orbits and evaluate each one by conducting a $\chi$$^2$ analysis for the declination and line of sight velocity using the locations of the globular clusters, PAndAS-04,  PAndAS-09, PAndAS-10, PAndAS-11 and PAndAS-12 (see Table \ref{OM_table:1}) that lie along NW-K2.  We use these particular clusters as they were used by \cite{RefWorks:236} to determine the velocity gradient along NW-K2 so they provide a robust counterpart to the masks on NW-K1.  They are also the globular clusters used by \cite{RefWorks:313} in their test particle models and those found to be associated with NW-K2 by \cite{RefWorks:405} and \cite{M:1147}.

Table \ref{OM_table:30} and Figure \ref{OM_Fig_10} show the results of the $\chi^2$ analysis.  The middle and right hand panels, of Figure \ref{OM_Fig_10}, show the systemic velocity of PAndAS-12 as a function of ${\mu^*_{\alpha}}$ and the heliocentric distance of PAndAS-12 as a function of ${\mu^*_{\alpha}}$ respectively.  The left hand panel shows ${\mu_{\delta}}$ versus ${\mu^*_{\alpha}}$ and indicates that there are two possible solutions for the proper motions for NW-K2 i.e. one for the stream moving away from and one with the stream moving towards M31. With the highest density of low $\chi^2$ values appearing in this latter region of the plot this could be indicative of a preference for the motion of NW-K2 being towards M31, which is consistent with findings by \cite{RefWorks:236} and \cite{RefWorks:313}.
  
\subsection{Stream Fits} \label{OM_Stream_Fits}
\subsubsection{Fitting NW-K1} \label{OM_Fitting the NW-K1 Stream}

The analysis in Section \ref{OM_Modelling_NW-K1} provides a good indication of the orbit of NW-K1's potential progenitor, And XXVII.  However, a more accurate delineation of the orbit can be obtained by fitting the track of the stream.  We note that there are multiple ways of doing this:
\begin{itemize}
\item Fitting to the centres of the masks (see Table \ref{OM_table:1a}).
\item Fitting Gaussian and background models to determine coordinates at points along the track of the stream and then fitting the orbit to these coordinates.
\end{itemize}
For this latter analysis we first obtain the clearest view possible of the stream using the density plot of the photometric metallicities,  [Fe/H]$_{\rm phot}$, in the quadrant of the M31 halo where NW-K1 is located (see Figure \ref{OM_Fig1}).  We select data from the PAndAS catalogue that meets the following criteria:
\begin{itemize}
\item Objects must be point sources, i.e. most likely to be stars.
\item They must lie close to the And XXVII/NW-K1 red giant branch, i.e. with 20.5 $\le$ \textit{i$_0$} $\le$ 24.5. 
\item They must have photometric metallicities in the range $-$2.5 $\leq$ [Fe/H]$_{\rm phot}$  $\leq$ $-$0.5, following the approach taken by \cite{RefWorks:76}. 
\end{itemize}

\begin{table}
	\centering
	\setlength\extrarowheight{2pt}
	\caption[Stream Track Coordinates]
	{Derived coordinates for the centre of NW-K1 in tangent plane coordinates centred on M31.}		
	\label{OM_table:2}
	\begin{tabular}{ccc} 
		\hline
		Box & $\xi$($\circ$) & $\eta$($\circ$) \\
		\hline 
0   &   $-$0.2  &  3.22 $\pm$ 0.25    \\     
1   &   $-$0.6  &  3.83 $\pm$ 0.03  \\      
2   &   $-$1.0  &  4.20 $\pm$ 0.04  \\     
3   &   $-$1.4  &  4.78 $\pm$ 0.56  \\    
4   &   $-$1.8  &  4.77 $\pm$ 0.21 \\     
5   &   $-$2.2  &  5.21 $\pm$ 0.08  \\      
		\hline
	\end{tabular}
\end{table}

Next we find the location of central points along the stream.  We overlay six 0.4$^{\circ}$ wide boxes across the stream. The box lengths are set to $\sim$twice the box width so that we focus our analysis on stars very near the stream and avoid including the nearby dwarf galaxy \mbox{And XXV} in the final box. We fit Gaussians to the locations of the stars in the box and define the probability that the location of the centre of the stream in a given box is at $\eta$$_0$ as :
\begin{equation} 
\label{OM_eq:010}
P_{\rm {cent},j}(\eta_{\rm j}, \eta_{\rm{unc},j}) = \frac{1}{\sqrt{2\pi\eta_{\rm{unc},j}^2} }  \mathrm{exp}\Bigg[\frac{-(\eta_{\rm j} - \eta_0)^2}{2 \eta_{\rm{unc},j}^2} \Bigg] ,
\end{equation} 
where \textit{P}$_{\rm {cent},j}$ is the posterior distribution function (pdf), $\eta$$_{\rm j}$ is the $\eta$ coordinate of the stream in the box with an uncertainty of $\eta$$_{\rm{unc},j}$  and $\eta$$_0$ is the $\eta$ coordinate of a known location in the box, e.g. the centres of the masks.

\begin{figure*}
  	\centering
	\includegraphics[height=.32\paperheight, width=.87\paperwidth]{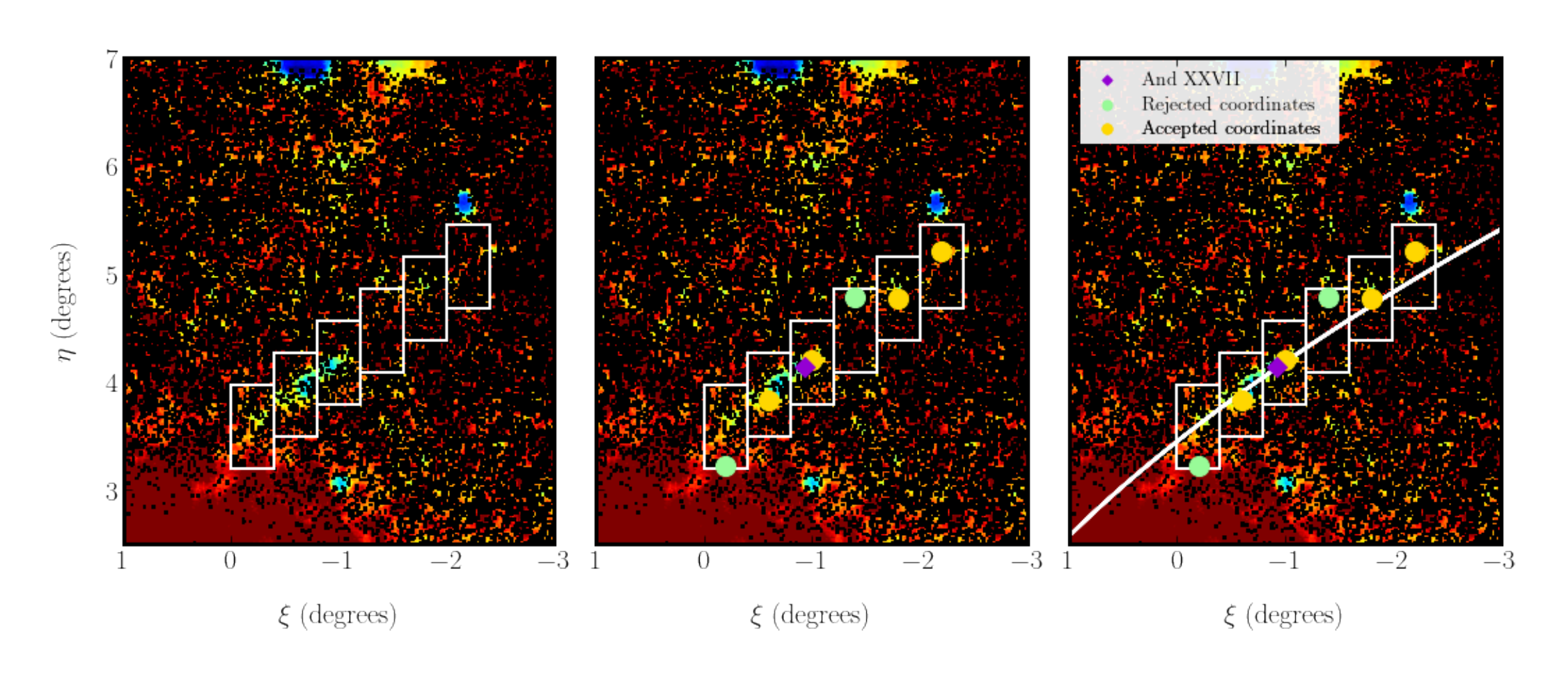}
    	\vspace*{-11mm}\caption[On-sky locations of the derived coordinates for the centre of the NW-K1 stream track]
	{On-sky locations of the derived coordinates for the centre of the NW-K1 stream track. The left hand panel shows the boxes (white open rectangles) overlaid onto the stream track.  On the middle and right hand panels the location of And XXVII is indicated by a purple diamond and the positions of the derived stream track coordinates. The gold circles indicate the coordinates used to determine potential best fit orbits, while the green circles indicate those discarded from further analysis.  On the right hand panel the stream coordinates are overlaid with an example of the best fit orbit, obtained from our MCMC analysis. 	 }
	\label{OM_Fig37}
\end{figure*}

We also assume that the background can vary linearly across each box so we describe the probability of a background star as : 
\begin{equation} 
\label{OM_eq:008}
P_{\rm bg, j} (T_{\rm U}) = \frac{T_{\rm L} + \bigg(\frac{T_{\rm U} - T_{\rm L} } {\eta_{\rm max}  -  \eta_{\rm min}} \bigg) (\eta_{\rm j}  - \eta_{\rm min})}{A} ,
\end{equation}
where: \textit{T}$_{\rm L}$ and \textit{T}$_{\rm U}$  are the lower and upper values for the background, $\eta$$_{\rm max}$ and $\eta$$_{\rm min}$ are the limits of the box, $\eta$$_{\rm j}$ is the $\eta$ coordinate of the stream track and \textit{A}  is the area under the background function, given by:

\begin{equation} 
\label{OM_eq:009}
A  = \bigg(\frac{T_{\rm U} + T_{\rm L} } {2} \bigg) ({\eta_{\rm max} - \eta_{\rm min})} ,
\end{equation}

The likelihood function for the location of the stream track can then be defined by:
\begin{equation} 
	\label{OM_eq:011}
	\begin{multlined}
		\mathcal{L}_{\rm bg}(\eta,  \eta_{\rm unc}, T_{\rm U}) = \sum_{i=1}^{\textit{N}}\mathrm{log}(nP_{\rm {cent},j}(\eta_{\rm j}, \eta_{\rm{unc},j}) + (1-\textit{n})\textit{P}_{\rm bg, j} (T_{\rm U})) ,
	\end{multlined}
\end{equation}
where: \textit{N} is the number of stars and \textit{n} is the fraction of the data within the stream.

We use the above equations in an MCMC analysis of our data using the {\sc emcee} software algorithm, \cite{RefWorks:571}, \cite{RefWorks:63}. We define the priors to be: 
\begin{itemize}
\item $\eta$$_{\rm min}$ < $\eta$ < $\eta$$_{\rm max}$
\item 0 < $\eta$$_{\rm{unc}}$ < width of the box
\item 0 < \textit{n} < 1 and \textit{T}$_{\rm U}$ > 0. 
\end{itemize}
Values for $\eta_j$, $\eta$$_{\rm{unc},j}$, \textit{T}$_{\rm U}$ and \textit{n} are initialised using a uniform distribution based on the priors, except for \textit{T}$_{\rm U}$ where the condition for the initial value is defined as 0 < \textit{T}$_{\rm U}$ < 10. We initialise \textit{T}$_{\rm L}$ to 1 but do not define priors as we do not fit this parameter.  We then let our Bayesian analysis run with 100 walkers taking 50,000 steps with a burn in of 10,000. To ensure that the chains have converged, we check the autocorrelation time ($\tau$) and find it to be in the range 50 < $\tau$ < 110.  This indicates that our number of steps is well above the recommended 10$\tau$ (\citealt{RefWorks:382}) and is sufficient to ensure a robust number of independent samples to allow our chains to converge and ensure the parameters are well constrained. Our results for the coordinates of the stream are shown in Table \ref{OM_table:2}

The left hand panel of Figure \ref{OM_Fig37} shows the positions of the boxes overlaid on NW-K1.  The middle panel shows the same information together with the derived coordinates for the centres of each box. We note that in the first box there is a considerable amount of noise from the M31 halo, so it is unlikely that this location is reliable.  Similarly for the fourth box, where there is a gap in the stream, the result also appears unreliable.  So we discard these two sets of coordinates and use the remaining four to find the best fit orbit for the stream track, an example of which is shown in the right hand panel of Figure \ref{OM_Fig37}.

We then use the same approach to fit these coordinates and the mask centres along with the velocity of the stream at these locations. As we do not have velocity data for the stream coordinates we use the systemic velocities from the mask centres, derived by P19 and listed in Table \ref{OM_table:1a}.  As there are no reliable distances to the derived coordinates or the masks, we set the initial value to the heliocentric distance of And XXVII.  We do not analyse distances along the stream nor do we use them in the fitting process. So our approach is not intended to produce an exact representation of NW-K1 but to model the track of the stream and the radial velocities along its length.

We use the same linear interpolation approach as described in section \ref{OM_Sim_Set_up} to obtain the values from the model orbit that are the closest match to the observed data.  We define the probability functions for the orbit matching the stream coordinates as:

\begin{equation} 
	\label{OM_eq:012}
		P_{\delta}(\eta_{\rm calc}, \eta_{\rm{unc}})  =  \frac{1}{\sqrt{2\pi\sigma_{\rm{unc}}^2} }  \mathrm{exp}\Bigg[\frac{-(\eta_0 - \eta_{\rm calc})^2}{2 \sigma_{\rm{unc}}^2} \Bigg] ,
\end{equation}

\begin{equation} 
	\label{OM_eq:013}
		P_{\rm vel}(v_{\rm mask}, v_{\rm{unc}})  =  \frac{1}{\sqrt{2\pi v_{\rm{unc}}^2} }  \mathrm{exp}\Bigg[\frac{-(v_0 - v_{\rm mask})^2}{2 v_{\rm{unc}}^2} \Bigg] ,
\end{equation}
where \textit{P}$_{\delta}$ and \textit{P}$_{\rm vel}$ are the pdfs; $\eta$$_{\rm 0}$ is the interpolated $\eta$ coordinate for the generated orbit; $\eta$$_{\rm calc}$ is the calculated stream coordinate with an uncertainty of $\eta$$_{\rm{unc}}$; \textit{v}$_0$ is the interpolated velocity coordinate for the generated orbit; \textit{v}$_{\rm mask}$ is the systemic velocity for the mask with an uncertainty of \textit{v}$_{\rm{unc}}$.

The likelihood function is given by:
\begin{equation} 
	\label{OM_eq:015}
	\begin{split}
		\mathcal{L}(\eta_{\rm calc}, \eta_{\rm{unc}}, v_{\rm mask}, v_{\rm{unc}}) = \sum_{i=1}^{\rm \textit{N}_{\rm c}}\mathrm{log}(P_{\delta}(\eta_{\rm calc}, \eta_{\rm{unc}}) + \\ \sum_{i=1}^{N_{\rm m}}\mathrm{log}(P_{\rm vel}(v_{\rm mask}, v_{\rm{unc}})) ,
	\end{split}
\end{equation}
where \textit{N}$_{\rm c}$ is the number of stream coordinates and \textit{N}$_{\rm m}$ is the number of spectroscopically observed fields.

As noted earlier, the $\chi$$^2$ analysis of the stream models shows that there are two possible solutions for the proper motion of the stream. This could be problematic for the {\sc emcee} software as it is not designed to handle separated multimodal posteriors where the likelihood is very low between the separate solutions.  So to ensure our chains can converge we fit the two solutions for the stream separately.  

First, we take the best fit orbit with motion in the direction towards M31. We initialise the walkers based on results from the best fit $\chi$$^2$ analysis.  We select a large range for the proper motions, constrained only by our expectation that the stream is unlikely to be moving faster than the escape velocity and centred around the proper motion for M31. 

We then run our Bayesian analysis using {\sc emcee} with 100 walkers taking 50,000 steps with a burn in of 10,000. We initialise the walkers in small distributions about the maximum likelihood, as recommended by \cite{RefWorks:382} and run separate models of the orbit using the stream coordinates and the mask centres using the priors shown in Table \ref{OM_table:22}.

\begin{table}
	\centering
	\setlength\extrarowheight{2pt}
	\caption[Stream Priors]
	{Priors for the MCMC analysis of NW-K1 for the directions of motion towards and away from M31}		
	\label{OM_table:22}
	\begin{tabular}{lcc} 
		\hline
Prior  & To   & Away \\
\hline 
${\mu^*_{\alpha}}$ /mas/yr     & 0.0 - 0.2                   & $-$0.2 - 0.0    \\
$\mu$$_{\delta}$ /masyr        & $-$0.2 - 0.0               & 0.0 - 0.2   \\
v$_{\rm A27} $                       & $-$580.0 - $-$500.0  &  $-$580.0 - $-$500.0 \\
Distance  (kpc)                       & 600.0 - 1000.0           &  600.0 - 1000.0 \\
		\hline
	\end{tabular}
\end{table}

 We then repeat the above process for the best fit orbit moving away from M31 again initialising the walkers based on results from the best fit $\chi$$^2$ analysis and running our Bayesian analysis as described above. 
\subsubsection{Fitting NW-K2} \label{OM_Fitting the NW-K2 Stream}

To obtain a best fit orbit along NW-K2 we use the centres of the observed masks associated with the stream as well as the co-ordinates of some of the co-located globular clusters as proxy for coordinates of the stream. We do not determine coordinates for the centre of the stream as results from the equivalent analysis of NW-K1 show a strong similarity between the model orbits derived from the centres of the observed masks and the stream coordinates.  As we adopted the observed data for further NW-K1 analysis, we decided that deriving stream coordinates for NW-K2 would only add complexity and not yield any useful insights.

We use the same linear interpolation approach as described in Section \ref{OM_Fitting the NW-K1 Stream} to obtain values for the model orbit that are the closest match to the observed data. We use the probability functions shown at Equations \ref{OM_eq:012} and \ref{OM_eq:013} and the likelihood function shown in Equation \ref{OM_eq:015} to model our data.  As with NW-K1, the $\chi$$^2$ analysis of NW-K2 indicated two possible solutions for the proper motion of the stream.  We note that \cite{RefWorks:405} determined that the simulations of the stream moving away from M31 (i.e. Case B as reported in \citealt{RefWorks:313}) were not viable, however, for completeness, we include both possible directions for NW-K2 and model each separately by setting the priors for the Bayesian analysis as shown in Table \ref{OM_table:22a}.

\begin{table}
	\centering
	\setlength\extrarowheight{2pt}
	\caption[Stream Priors]
	{Priors for the MCMC analysis of NW-K2 for the directions of motion towards and away from M31}		
	\label{OM_table:22a}
	\begin{tabular}{lcc} 
		\hline
Prior  & To   & Away \\
\hline 
${\mu^*_{\alpha}}$ /mas/yr     & 0.0 - 0.2                       & $-$0.2 - 0.0    \\
$\mu$$_{\delta}$ /masyr        & $-$0.2 - 0.0                  & 0.0 - 0.2   \\
v$_{\rm A27} $                       & $-$600.0 - $-$400.0     &  $-$600.0 - $-$400.0 \\
Distance  (kpc)                       & 600.0 - 1000.0             &  600.0 - 1000.0 \\
		\hline
	\end{tabular}
\end{table}

In both cases we run our Bayesian analysis as described for NW-K1 and again, to ensure that in all our runs the chains have converged, we check the autocorrelation time ($\tau$) finding it to be within the acceptable range.  

\subsection{Orientation Relative to M31} \label{OM_Orientation}

To provide further confirmation of the directions of the streams with respect to M31, we derive the 3-d velocity vectors for their best fit orbits using:
\begin{equation} 
\label{OM_eq:009}
\boldsymbol{\vec{v_{3d}}} = (\boldsymbol{\vec{r}} \cdot \boldsymbol{\vec{v})}/ \lVert r \rVert
\end{equation}
where: $\boldsymbol{\vec{v_{3d}}}$ is the 3-d velocity vector for the stream/direction, $\boldsymbol{\vec{r}}$ is the position vector for the stream/direction along the track and $\boldsymbol{\vec{v}}$ is the corresponding velocity vector.  We find the derived values for the streams moving away from M31 to be positive and those for the streams moving towards M31 to be negative, indicating that the streams are, indeed, moving in the directions we have described.

\subsection{Increased Orbital Integration Time} \label{OM_Increased_Orbit}

As part of our analysis we also integrate samples from our posterior chains over a 5 Gyr timeframe, enabling us to:
 \begin{itemize}
\item Determine if there is any connection between NW-K1 and NW-K2 in the distant past or for different mass models of M31.
\item Calculate the perigalacticons for NW-K1 and NW-K2.
\item Derive the tidal radius for And XXVII to provide further insights into whether or not this dwarf spheroidal galaxy is in the process of being disrupted.
\item Ascertain if there is any possible association between NW-K2 and the dwarf elliptical galaxies NGC 147 and NGC 185, as suggested in \cite{RefWorks:516} and \cite{RefWorks:449}.
\end{itemize}

To derive the perigalacticons of the streams, we use the 100 random orbits for NW-K1 and NW-K2 and integrate each of them over 5 Gyrs.  We find the pericentres of each orbit and take the mean and standard deviation of these values to obtain an overall estimate for the pericentric radii of the stream tracks. 

To determine the tidal radius, and hence the strength of M31's tidal field on And XXVII, we use the rotation curve developed by \cite{M:1068} to obtain a value of 237.4 $\pm$ 7.8\kms {} for the circular velocity of M31 at And XXVII's pericentric radius, which enables us to determine the enclosed mass of M31 at this distance (\textit{M}$_{\rm M31}$) using:
\begin{equation} 
\label{OM_eq:010}
M_{\rm M31} = \frac{v_{\rm cM31}^2   r_{\rm peri}}{G}  ,
\end{equation}
where: \textit{v}$_{\rm cM31}$ is the circular velocity of M31 at the pericentric distance, \textit{r}$_{\rm peri}$ is the pericentric distance and G is the gravitational constant.

Then, following \cite{M:1023}, we determine the tidal radius for And XXVII with:
\begin{equation} 
\label{OM_eq:011}
r_{\rm tidal} = r_{\rm peri} {\bigg( \frac{M_{\rm A27} }{2M_{\rm M31}} \bigg)}^\frac{1}{3} ,
\end{equation}
where: \textit{r}${\rm _{\rm tidal} }$ is the tidal radius for And XXVII,  \textit{M}$_{\rm A27}$ is the enclosed mass of And XXVII at the half-light radius = 8.3$^{+2.8}_{-2.9}$ $\times$ 10$^7$\Msun {} (\citealt{RefWorks:42}) and the 2 in the denominator acknowledges the assumption that M31 has a flat rotation curve.  

\subsection{Variations of the M31 potential} \label{OM_M31_Potential}
Conscious that our analysis is based on a single mass model for the M31 potential, we re-run our orbital models using values for a heavier and a lighter potential for M31 for both streams in the direction towards M31.   We retain the same values for the Hernquist Bulge and Miyamoto-Nagai disk as described in Section \ref{OM_Stream Models}  and use the values shown in Table \ref{OM_table:5} for the lighter and heavier NFW halos for M31.  The halo concentrations are obtained via look-up using Figure 3 in \cite{M:1124}.  The scale radii for the two new potentials are then derived using equations \ref{OM_eq:027},  \ref{OM_eq:028} and \ref{OM_eq:029}.

\begin{table}
	\centering
	\setlength\extrarowheight{2pt}
	\caption[M31 potential variations]
	{Variations for M31 potential.  The lighter halo mass is obtained from \cite{M:999} and the heavier from \cite{RefWorks:378}.  The halo concentrations are obtained via look-up using Figure 3 in \cite{M:1124}.  The scale radii for the two new potentials are then derived using equations \ref{OM_eq:027},  \ref{OM_eq:028} and \ref{OM_eq:029}.
	}	
	\label{OM_table:5}
	\begin{tabular}{lccc} 
		\hline
		  & Halo Mass                                 &  Scale radius    & Halo  \\
		   & {\Msun}  $\times$  10$^{12} $  &  kpc                 & concentration \\
		\hline 
Lighter    & 1.33    &  23.2  & 10 \\
Main       &  1.995 &  28.9  & 8.9\\
Heavier  &  2.85   &  29.9  & 10 \\
		\hline
	\end{tabular}
\end{table}

\begin{equation} 
\label{OM_eq:027}
\rho_c = 3 H_0^2 / 8 \pi G, 
\end{equation}

\begin{equation} 
\label{OM_eq:028}
r_{200}^3 = 3 M_{200} / 200 \times 4 \pi \rho_c, 
\end{equation}

\begin{equation} 
\label{OM_eq:029}
r_s = r_{200} / c_{200}, 
\end{equation}

\noindent where $\rho_c$ is the density (kg/m$^3$), H$_0$ is the Hubble constant (\kms Mpc), G is the gravitational constant (Nm$^2$/kg$^2$), M$_{200}$ (\Msun) is the mass of the NFW halo, r$_{200}$ (kpc) is the critical radius, r$_s$ (kpc) is the scale radius and c$_{200}$ is the concentration parameter.

\subsection{Integrals of Motion Analysis} \label{OM_Integrals of Motion}

\begin{figure}
	\includegraphics[height=.3\paperheight, width=\columnwidth]{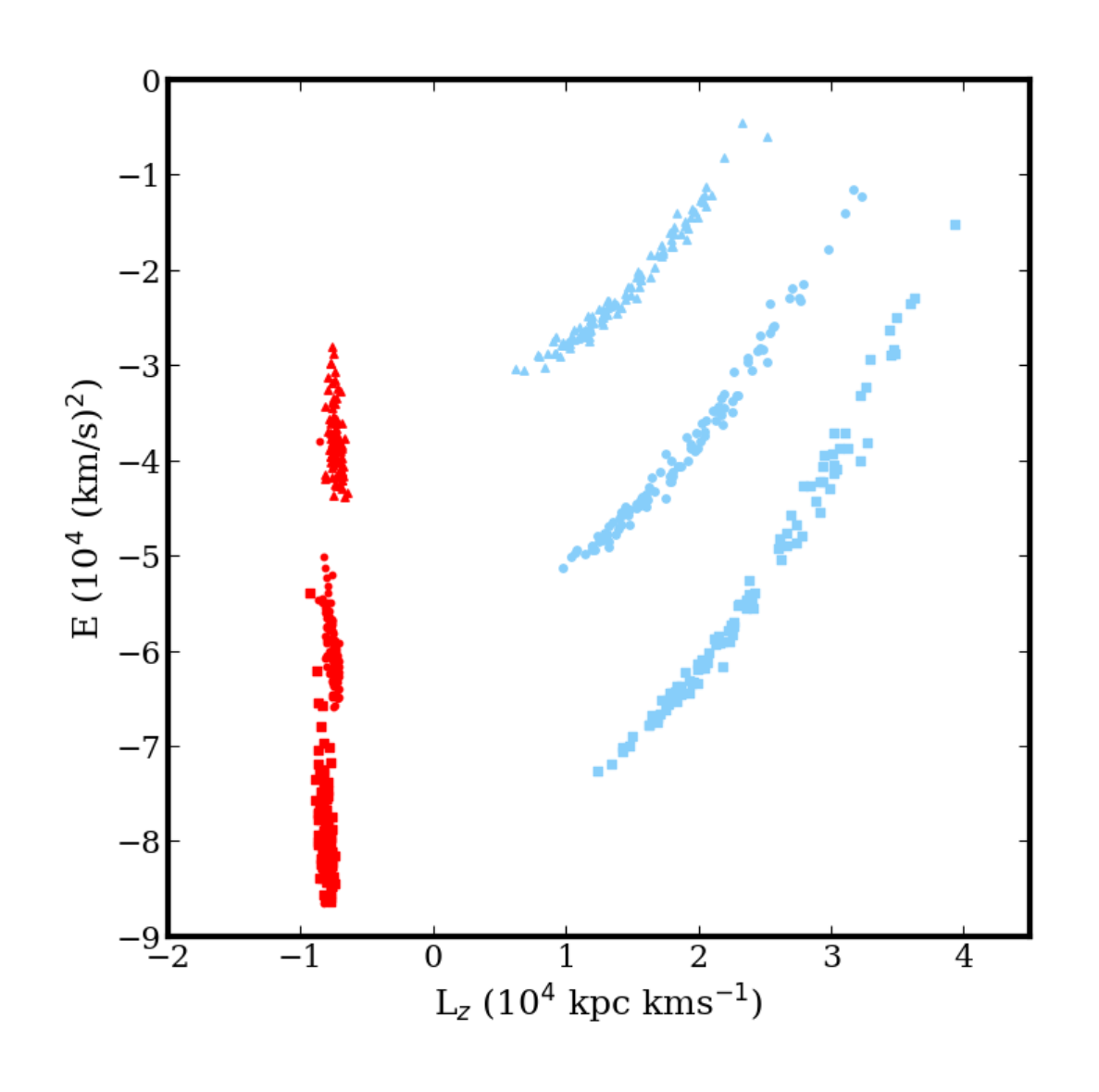}
    	\vspace*{-9mm}\caption[IoM analysis of streams]
	{L$_z$vs E for 100 random orbits for NW-K1 (light blue) and NW-K2 (red) for the main (circles), heavier (squares) and lighter (triangles) M31 potentials. 
	}
	\label{OM_Fig180}
\end{figure}

\cite{M:830} reported that stream trajectories could be characterised by the Integrals of Motion (IoM) elements energy (E) and the z-component of angular momentum (L$_z$).  Several groups have used these parameters to: identify new streams and new stream members (e.g. \cite{M:1141}, \cite{M:1143}, \cite{M:1142}, \cite{M:1053}); explore the properties of streams to determine if they are separate entities or share a progenitor (e.g. \cite{RefWorks:746}, \cite{M:1145}) and associate streams with potential progenitors (e.g. \cite{M:1146}).
 We undertake an IoM analysis of our streams, under the influence of our various M31 potentials, using:
 
\begin{equation} 
\label{OM_eq:120}
E = 0.5*(v_x^2 + v_y^2 + v_z^2) + \phi_{\rm M31}(r)
\end{equation}

\noindent where \textit{v}$_x$, \textit{v}$_y$, \textit{v}$_z$ are the components of the velocity vector for the orbit and $\phi$$_{\rm M31}$(\textit{r}) are the potentials for M31 shown in Table \ref{OM_table:5}.  E is determined per unit mass, hence there is no mass parameter in the kinetic energy element of the equation.  We derive the angular momentum using: 
\begin{eqnarray}
\label{OM_eq:121}
L_x = yv_z - zv_y \\
L_y = zv_x - xv_z \\
L_z = xv_y - yv_x
\end{eqnarray}

\noindent where: \textit{L}$_x$, \textit{L}$_y$ and \textit{L}$_z$ are components of angular momentum; \textit{x}, \textit{y} and \textit{z} are the cartesian coordinates of And XXVII (for NW-K1) or PAndAS-12 (for NW-K2) and \textit{v}$_x$, \textit{v}$_y$, \textit{v}$_z$ are components of the stream velocity vectors.  We present our results in Figure \ref{OM_Fig180}.

\begin{figure*}
  	\centering
	\includegraphics[height=.3\paperheight, width=.8\paperwidth]{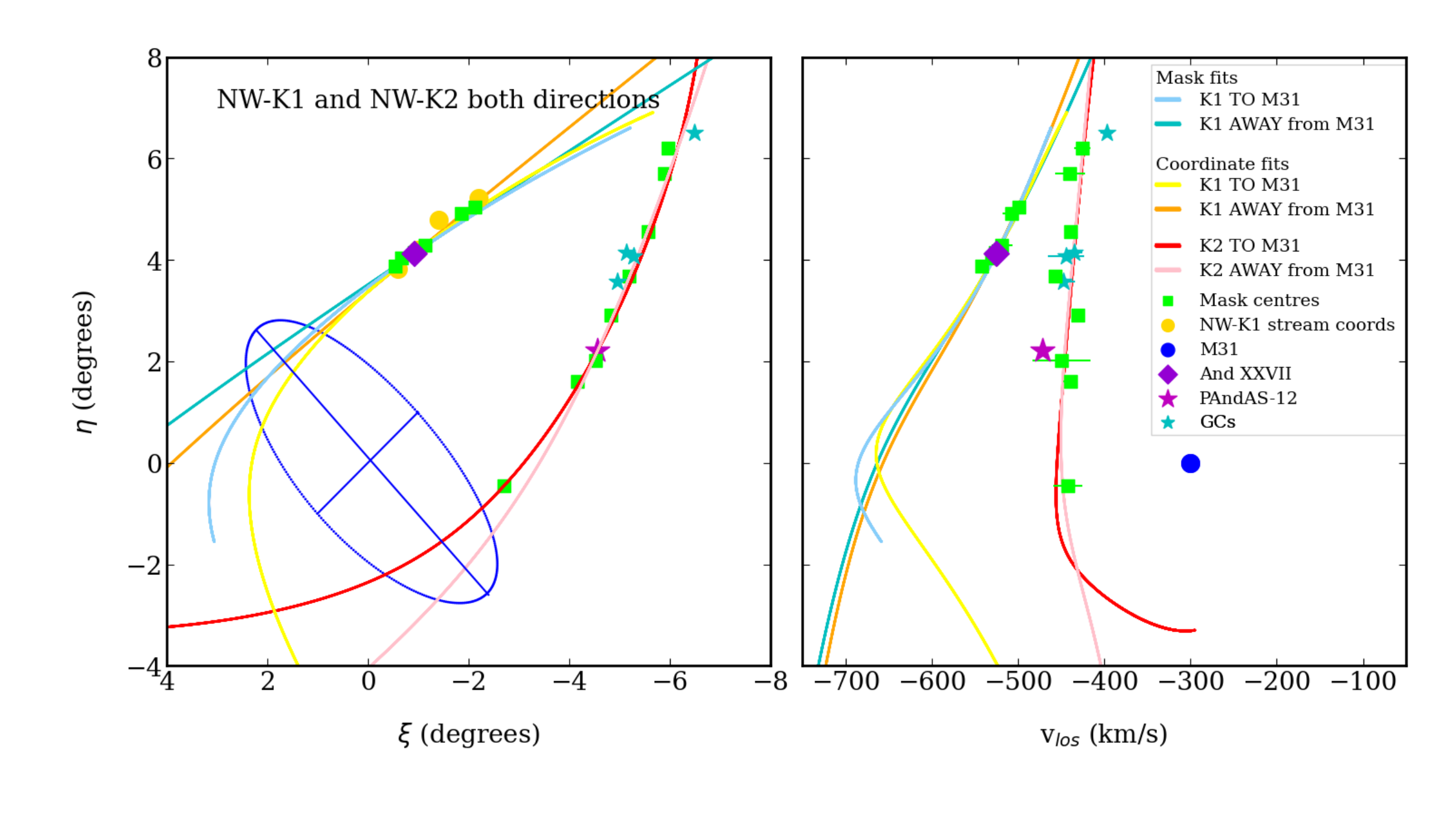}
    	\vspace*{-6mm}\caption[Best fit orbits for NW-K1]
	{Best fit orbits for NW-K1and NW-K2 obtained using the posterior values from the MCMC fitting processes.
	In both panels 
	the orange line represents the track obtained by fitting the stream coordinates and the cyan line represents the track obtained by fitting the mask centres for NW-K1 moving away from M31; 
	the yellow line represents the track obtained by fitting the stream coordinates and the light blue line shows the track obtained from fitting the mask centres for NW-K1 moving toward M31
	The tracks for NW-K2 are represented by pink line and red lines showing the model orbits moving away from towards M31, respectively.
	The left hand panel also shows the model orbits overlaid with the positions of And XXVII (purple diamond), the stream coordinates (gold circles), the mask centres (green rectangles), the GCs associated with NW-K2 (cyan stars) and PAndAS-12 (magenta star).
	The right hand panel shows the model orbits overlaid with the systemic velocities of the mask centres (green rectangles), And XXVII (purple diamond) the GCs associated with NW-K2 (cyan stars), PAndAS-12 (magenta star) and M31 (blue circle).
  The blue ellipse traces the M31 halo, taking a semi-major axis of  55 kpc with a flattening of 0.6, \cite{RefWorks:107}.	 
  }
	\label{OM_Fig18}
\end{figure*}

\section{Results and Discussion} \label{OM_Discussion} 

In this section we present the results arising from our modelling and fitting orbits for the two components, NW-K1 and NW-K2, of the NW Stream. 
\subsection{Stream direction} \label{Stream direction}  
Figure \ref{OM_Fig18} shows the best fit orbits for NW-K1 and NW-K2 generated using posteriors from the datasets and processes described in Sections \ref{OM_Fitting the NW-K1 Stream} and \ref{OM_Fitting the NW-K2 Stream}.  The plots indicate that the models are good representations of the data.  

For both NW-K1 and NW-K2 we see that the tracks moving away from M31 are virtually straight lines, indicative of unbound orbits, whereas there are indications of possible pericentres along the orbits for the tracks moving towards M31. We also note that the track of NW-K2 (shown in the lefthand panel) and the plot of the line of sight radial velocities along the stream (righthand panel) are consistent with results obtained by \cite{RefWorks:313} and are also supportive of the hypothesis that NW-K2 is moving towards M31. 

\begin{figure*}
  	\centering
	\includegraphics[height=.80\paperheight, width=.70\paperwidth]{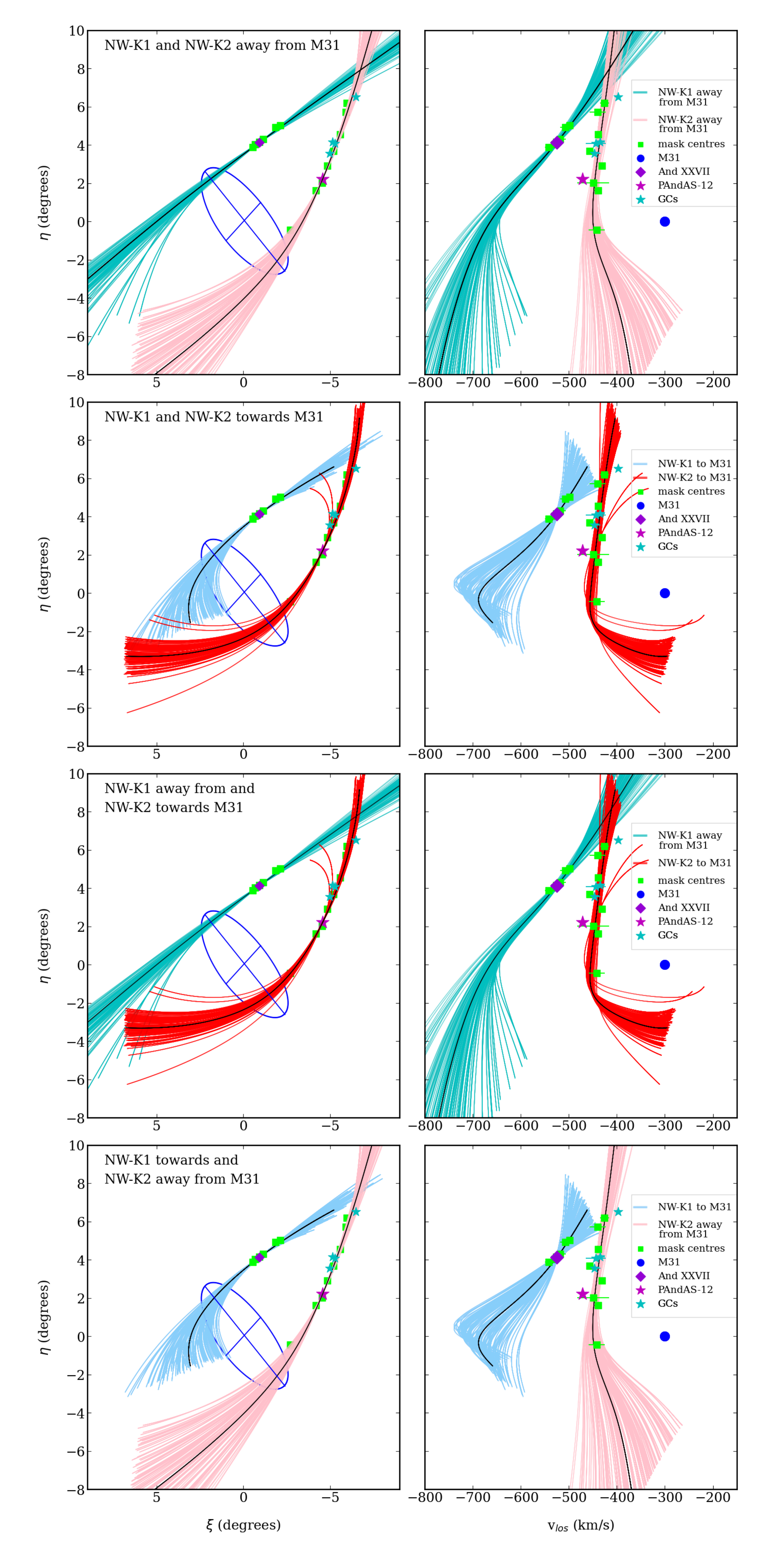}
    	\vspace*{-6mm}\caption[Line of sight velocities along model orbits for NW-K1 and NW-K2 moving towards M31]
	{Stream models for NW-K1 and NW-K2. The blue ellipse traces the M31 halo (using the same properties as Figure \ref{OM_Fig18}).  The green rectangles show the locations of the mask centres along both NW-K1 and NW-K2, the purple diamond indicates the position of And XXVII, the magenta star represents the GC PAndAS-12 with the other GCs represented by cyan stars.  In each set of panels the left hand plot shows the on-sky track of the potential orbits and the right hand plot shows the associated line of sight velocities.  In all panels, the best fit orbit is denoted by a black line. 
	The top two panels show 100 random orbits for NW-K1 (cyan lines) and NW-K2 (pink lines) away from M31.  
	The next two panels show 100 random orbits for NW-K1 (light blue lines) and NW-K2 (red lines) moving towards M31.
	The next two panels show 100 random orbits for NW-K1 (cyan lines) moving away from M31 and NW-K2 (red lines) moving towards M31.
	The bottom two panels show 100 random orbits for NW-K1 (light blue lines) moving towards M31 and NW-K2 (pink lines) moving away from M31.
	}
	\label{OM_Fig24}
\end{figure*}

To test if there are any other viable orbits that could connect the two streams, we obtain 100 randomly selected orbits from the posterior chains of our stream fitting process of NW-K1 and NW-K2 and plot them see Figure \ref{OM_Fig24}.  For simplicity and clarity we use the posterior chains produced from the analysis of the mask centres as proxy for all the NW-K1 models. 

Examining the relative speeds of the streams, using the magnitude of the 3-d velocity vector, we find that those moving away from M31 are $\sim$690 \kms{ }for NW-K1 and $\sim$349 \kms{ }for NW-K2. The large value for NW-K1 is indicative of an unbound orbit, which is consistent with the appearance of the stream tracks in the plots, while the value for NW-K2 is more in-keeping with a bound orbit. For the streams moving towards M31 we obtain relative speeds of \mbox{$\sim$250 \kms} for NW-K1 and $\sim$270 \kms{ }for NW-K2 which are consistent with bound orbits around M31.  These results, taken in conjunction with orientations of the 3-d velocity vectors enable us to discard the possibility that either stream is moving away from M31 and to conclude, in agreement with \cite{RefWorks:236}, \cite{RefWorks:313}, \cite{RefWorks:405} and P19, that both streams are most likely to be moving towards M31 on infall trajectories.

\subsection{Stream Structure} \label{Stream Structure}
  
\begin{figure*}
	\includegraphics[height=.30\paperheight, width=.85\paperwidth]{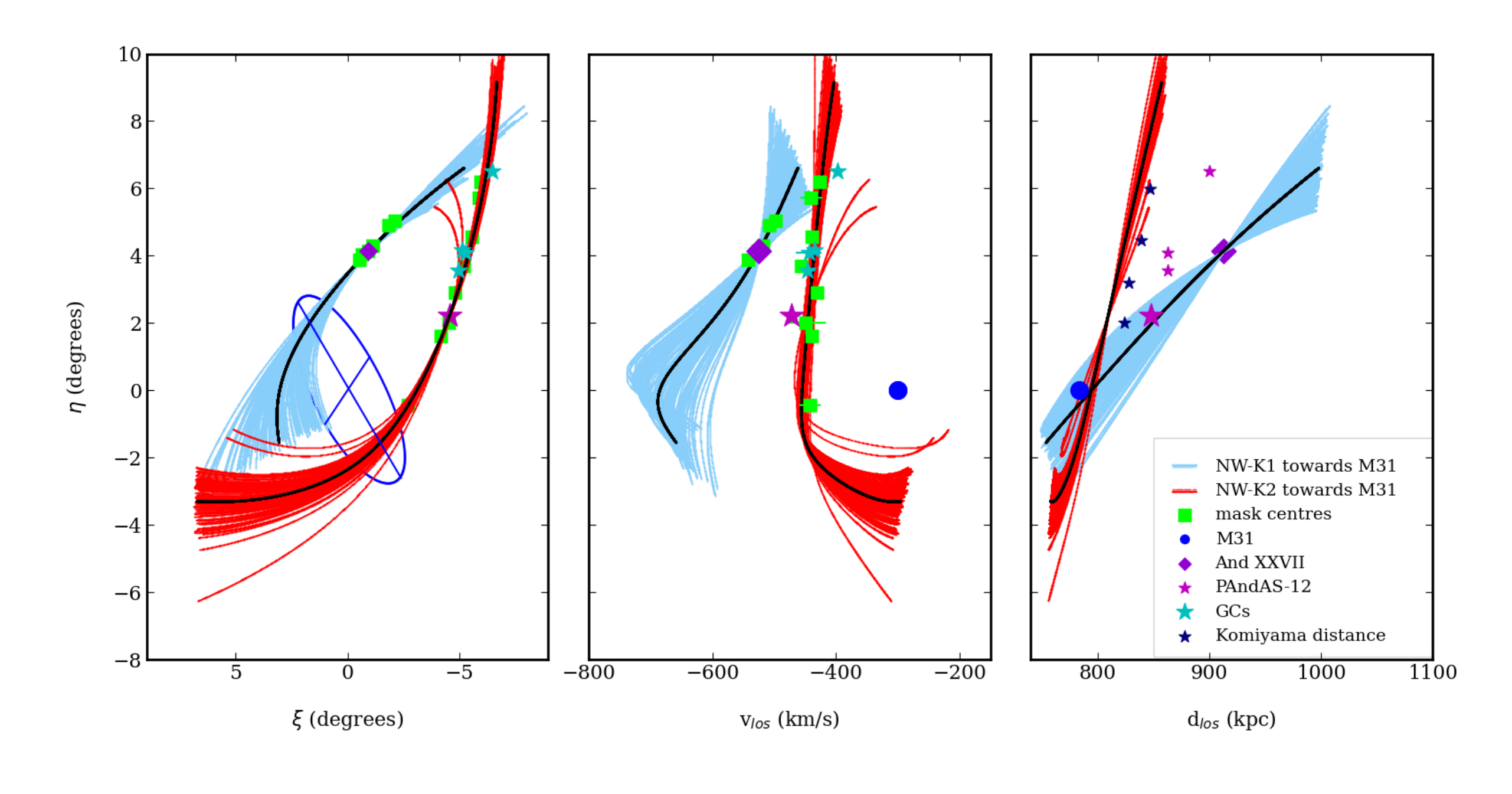}
    	\vspace*{-5mm}\caption[Line of sight velocities along model orbits for NW-K1 and NW-K2 as a function of on-sky location]
	{Line of sight velocities and distances along model orbits for NW-K1 and NW-K2 as a function of on-sky location. The leftmost panel reprises the on-sky locations of the orbits overlaid with the position of And XXVII (purple diamond),  the mask centres for both NW-K1 and NW-K2 (green rectangles), PAndAS-12 (magenta star), the other GCs associated with NW-K2 (cyan stars) and the M31 halo (blue ellipse).  The light blue lines show NW-K1 moving towards M31 and the red lines show the tracks of NW-K2 moving towards M31.  The middle panel shows the line of sight velocities along the stream tracks (colour coding the same as for the left hand panel) over-plotted with the systemic velocities for the masks (green rectangles), And XXVII (purple diamond), PAndAS-12 (magenta star) and the other GCs (cyan stars).   
The right hand panel shows the line of sight distances along the stream tracks for NW-K1 and NW-K2 (colour coding the same as for the left hand panel), over-plotted with the heliocentric distances of And XXVII (purple diamond), M31 (blue circle), PAndAS-12 and the other GCs (magenta stars) and distances to sections of the stream as calculated by \cite{RefWorks:405}.
}
	\label{OM_Fig26}
\end{figure*}

\begin{figure*}
	\includegraphics[height=.30\paperheight, width=.85\paperwidth]{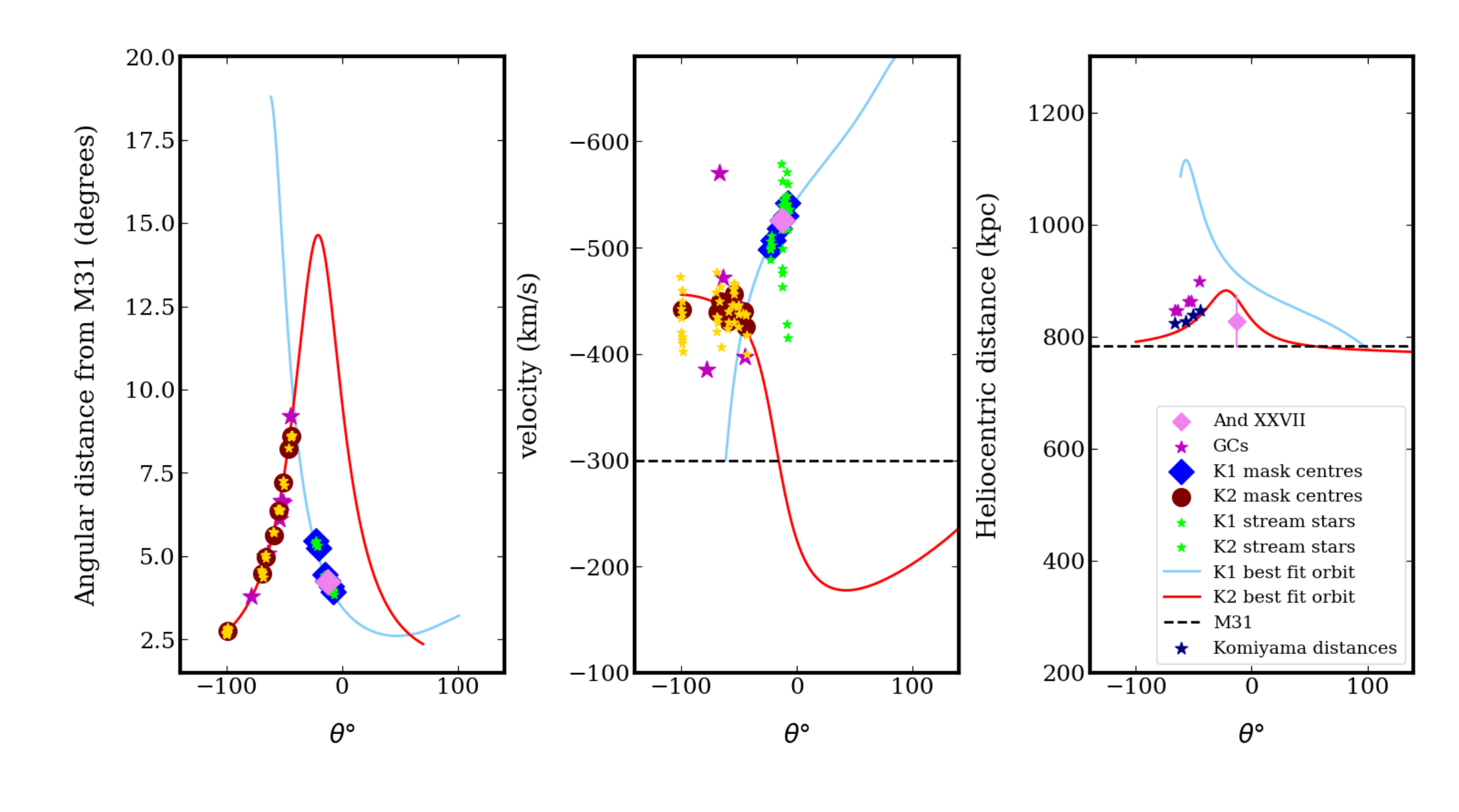}
    	\vspace*{-5mm}\caption[Line of sight velocities along model orbits for NW-K1 and NW-K2 as a function of on-sky location]
	{On-sky properties of NW-K1 and NW-K2 observable data overlaid with best fit orbits integrated over 3.5 Gyrs. The left hand panel shows the position of the observables (i.e. centres of the masks and stream stars thereon, GCs and And XXVII) as a function of distance from the centre of M31. The middle panel shows the systemic velocities of the observables as a function of distance from the centre of M31.  The right hand panel shows the heliocentric distances of the orbits, along with those of And XXVII, the GCs and distances to sections of the stream as calculated by \cite{RefWorks:405}.
}
	\label{OM_Fig53}
\end{figure*}

To explore the possibilities of NW-K1 and NW-K2 being separate streams or elements of a single structure we review their on-sky locations, radial velocities and heliocentric distances.  In Figure \ref{OM_Fig26}, with the best fit orbits integrated over 0.5 Gyrs,  we see that none of the plots are indicative of any connection between the two streams .  In Figure \ref{OM_Fig53}, which shows the best fit orbits integrated over 3.5 Gyrs, we find:

\begin{itemize}
\item Left hand panel: the plot does not provide confirmation of any particular stream morphology, though it does indicate that the orbit for a single structure would need to have a sharp turning point in order to pass through both sets of data from the observables; 
 
\item Middle panel: this plot has more of an indication that the two streams are likely to be separate structures due to their velocities being in the same area of the plot (i.e. are more negative) relative to the heliocentric velocity of M31 ($-$300 $\pm$ 4 km/s).  If the two streams were part of the same structure, we would expect to see most or all of the velocities from one of the streams lying below that of M31 which would indicate an orbit looping around M31. We do see that as the NW-K2 orbit loops down towards the NW-K1 stream (see left hand panel), however, once this happens, we note that its radial velocities have flipped with respect to M31 and are not aligned with those of NW-K1;

\item Right hand panel - here we note that both streams appear to lie behind M31 along all or most of their full lengths.  For NW-K2 this is in-keeping with work undertaken by Mackey et al (2015, private communication) 
who, based on observations described in \cite{RefWorks:236}, determined the distance moduli for some of the GCs associated with the stream (see Table \ref{OM_table:111}). It is also consistent with \cite{RefWorks:405}, who determined the distance moduli to four locations along the stream, reporting them to be 24.64 $\pm$ 0.20, 24.62 $\pm$ 0.18, 24.59 $\pm${ }0.18 and 24.58 $\pm$ 0.19.
\end{itemize}

\begin{figure*}
  	\centering
	\includegraphics[height=.40\paperheight, width=.75\paperwidth]{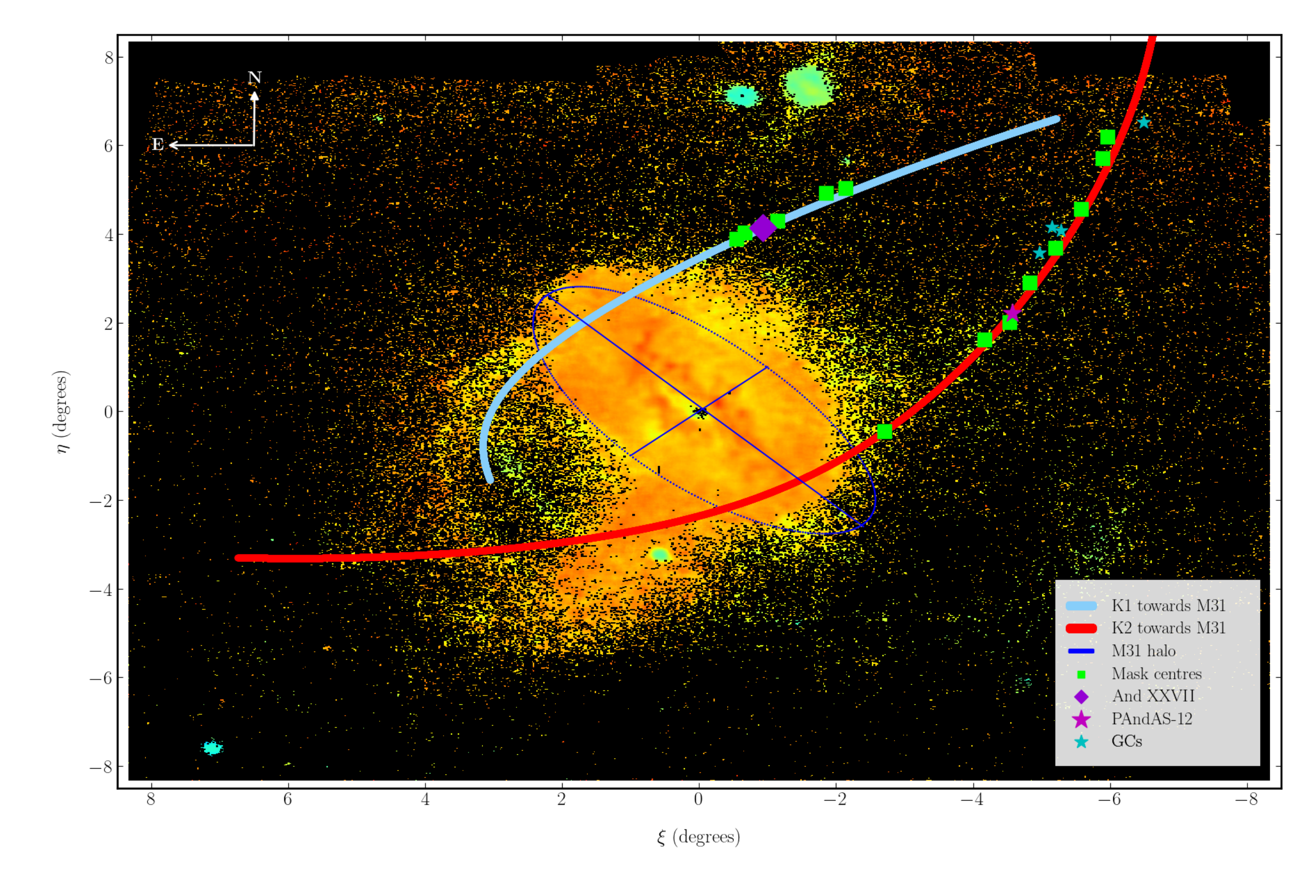}
    	\vspace*{-7mm}\caption[Best fit orbits for NW-K1 and NW-K2 overlaid on the PAndAS map]
	{Best fit orbits for NW-K1 and NW-K2 overlaid on a photometric map of M31. The light blue line represents the best fit model orbit for NW-K1 moving towards M31.  The red line shows the best fit model orbit for NW-K2 moving towards M31.
	The purple diamond indicates the location of And XXVII, PAndAS-12 is represented by a magenta star, all other GCs (PAndAS 04, 09, 10 and 11) are represented by cyan stars and the mask centres on both streams are represented by green rectangles.   The data includes point source objects from the PAndAS catalogue with $-$3.0 < [Fe/H] < 0.0.  The plot indicates that these two stream tracks do not connect to form a single structure. }
	\label{OM_Fig19}
\end{figure*}

\begin{figure}
  	\centering
	\includegraphics[height=.25\paperheight, width=\columnwidth]{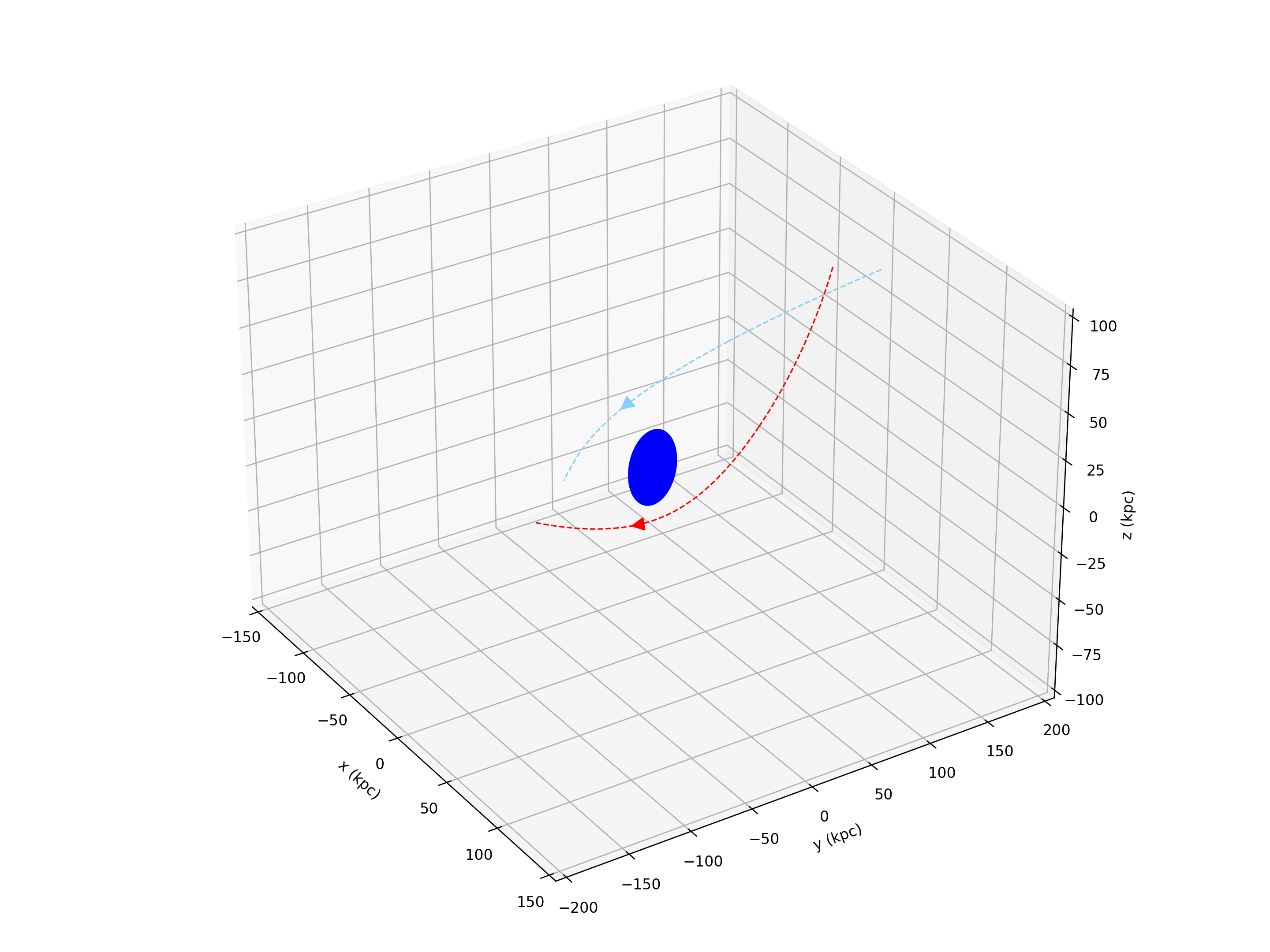}
    	\vspace*{-6mm}\caption[Frame from orbits movie]
	{3D projection of the NW-K1 and NW-K2 orbits (colour coded as in previous figures), both moving in a direction towards M31(represented by the ellipse).  A version of the full movie can be found at \url{https://youtu.be/UOO9s4DLgaU}.
	 }
	\label{OM_Fig28}
\end{figure}

\begin{figure*}
	\includegraphics[height=.80\paperheight, width=.8\paperwidth]{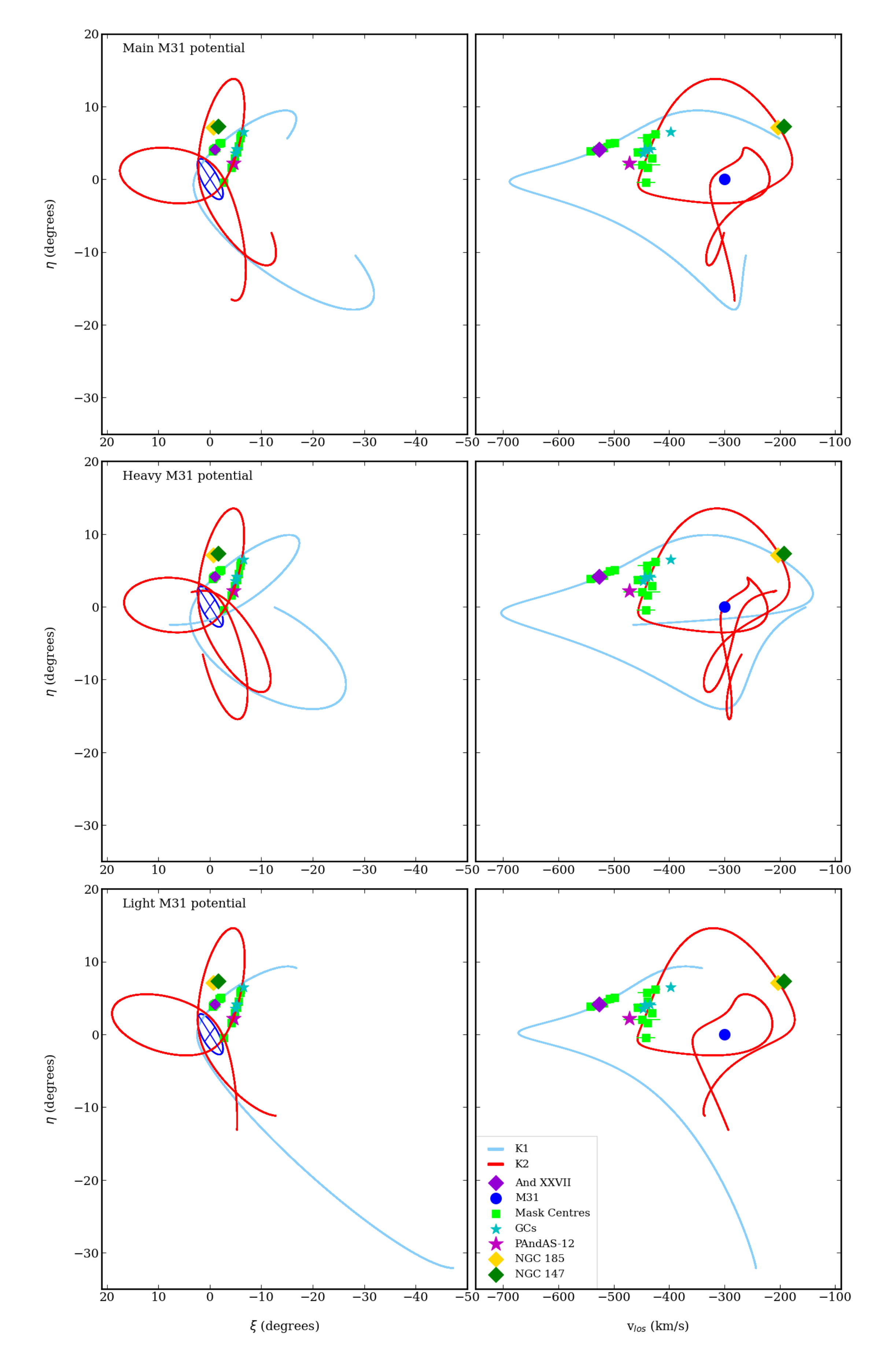}
    	\vspace*{-7mm}\caption[Model orbits for NW-K1 and NW-K2 generated over a 5 Gyr period.]
	{Model orbits for NW-K1 and NW-K2 generated over a 5 Gyr period, with varying potentials for M31, to determine whether they may be connected to each other further in the past.  The plots show both NW-K1 and NW-K2 streams with motion towards M31. The left hand plot shows the on-sky track of the potential orbits, together with the M31 halo (ellipse) and the right hand plot shows the associated line-of-sight velocities with that for M31 indicated by the circle. Both plots are overlaid with the positions and velocities of the mask centres for both NW-K1 and NW-K2, PAndAS-12, the other GCs associated with NW-K2, NGC 147, and NGC 185.  There does not appear to be a connection between the two streams over this longer timeframe or for the different model masses for the M31 potential. \\ 
	}
	\label{OM_Fig30}
\end{figure*}

In Figure \ref{OM_Fig19} we project the 0.5 Gyrs best fit orbits  onto a photometric map of M31. We see that they follow the tracks of their respective streams and are co-located with the centre of And XXVII and the centres of the masks along NW-K1 and with the GCs and mask centres along NW-K2, but do not appear to connect with one another.  Nor do they in a movie of the orbits, (Figure \ref{OM_Fig28} shows an example frame from the movie) or in the plots showing the best fit orbits integrated over 5 Gyrs with varying M31 potentials,  see Figure \ref{OM_Fig30} . Given the lack of convergence of the streams in these and the preceding plots, our results would appear to support the hypothesis that NW-K1 and NW-K2 are not part of a single structure and are more likely to be separate streams.

\begin{table}
	\centering
	\setlength\extrarowheight{2pt}
	\caption[Distance moduli and projected radii for NW-K2 GCs]
	{Distance moduli and projected radii for NW-K2 GCs from Mackey et al (2015, private communication) }		
	\label{OM_table:111}
	\begin{tabular}{lccc} 
		\hline
		Name & (m-M)$_0$ & $R_p$  \\ [0.5ex]
		           &                    & kpc\\ [0.5ex]
		\hline 
		PAndAS-04   & 24.77  & $\sim$125      \\ [0.5ex] 
		PAndAS-09   & 24.68  & $\sim$85 - 90    \\[0.5ex] 
		PAndAS-11   & 24.68  & $\sim$85 - 90  \\[0.5ex] 
		PAndAS-12   & 24.64  & $\sim$65 - 70    \\[0.5ex] 
		PAndAS-13   & 24.64  & $\sim$65 - 70   \\[0.5ex] 
		\hline
	\end{tabular}
\end{table}

\subsection{Stream Properties} \label{Stream Properties} 

In modelling and fitting the stream tracks for NW-K1 and NW-K2 we also derive their proper motions. When we compare these values to those from other works (see Table \ref{OM_table:4}) we see that our results for NW-K2 and for NW-K1 moving towards M31 are of a consistent order of magnitude. 

From our analysis of 100 random orbits integrated over a 5 Gyr timeframe, we obtain a pericentric distance of 28.7 $\pm$ 2.0 kpc for NW-K2, which aligns with the views of \cite{RefWorks:313} and \cite{RefWorks:405} that orbit for the progenitor of this stream would require a pericentre $\ge$ 25 kpc.  For NW-K1 we determine the pericentre to lie 42.9 $\pm$ 14.0 kpc from M31.  This is not consistent with estimates from \cite{M:1021}, who determined the pericentric radius for And XXVII to be 69 $\pm$10 kpc, albeit based on a much larger heliocentric distance for And XXVII than used in this work. 

We also derive a value of \mbox{1.8 $\pm$ 0.4 kpc} for the tidal radius of And XXVII. In comparison with the half-light radius of And XXVII (657$^{+112}_{-271}$ pc,  \citealt{RefWorks:42}) it is not immediately obvious that the galaxy is being tidally disrupted.  However, as \cite{RefWorks:206} experienced difficulties in obtaining a half-light radius for And XXVII, they concluded that it was nearing the end of its tidal disruption and was no longer a bound system.  Based on these findings, it is plausible that And XXVII may already have had most of its stars stripped away and it is entirely possible that the its half-light radius is significantly smaller today than it may have been during its approach to its pericentre. 

Our analysis of the streams in the IoM space, see Figure \ref{OM_Fig180}, shows that, for all three M31 potentials, NW-K1 has higher energy values for most of its orbits, while its angular momenta appear to be opposite those of NW-K2.  This would seem to support the hypothesis that NW-K1 and NW-K2 are separate streams.
 
\begin{figure*}
	\includegraphics[height=.25\paperheight, width=.8\paperwidth]{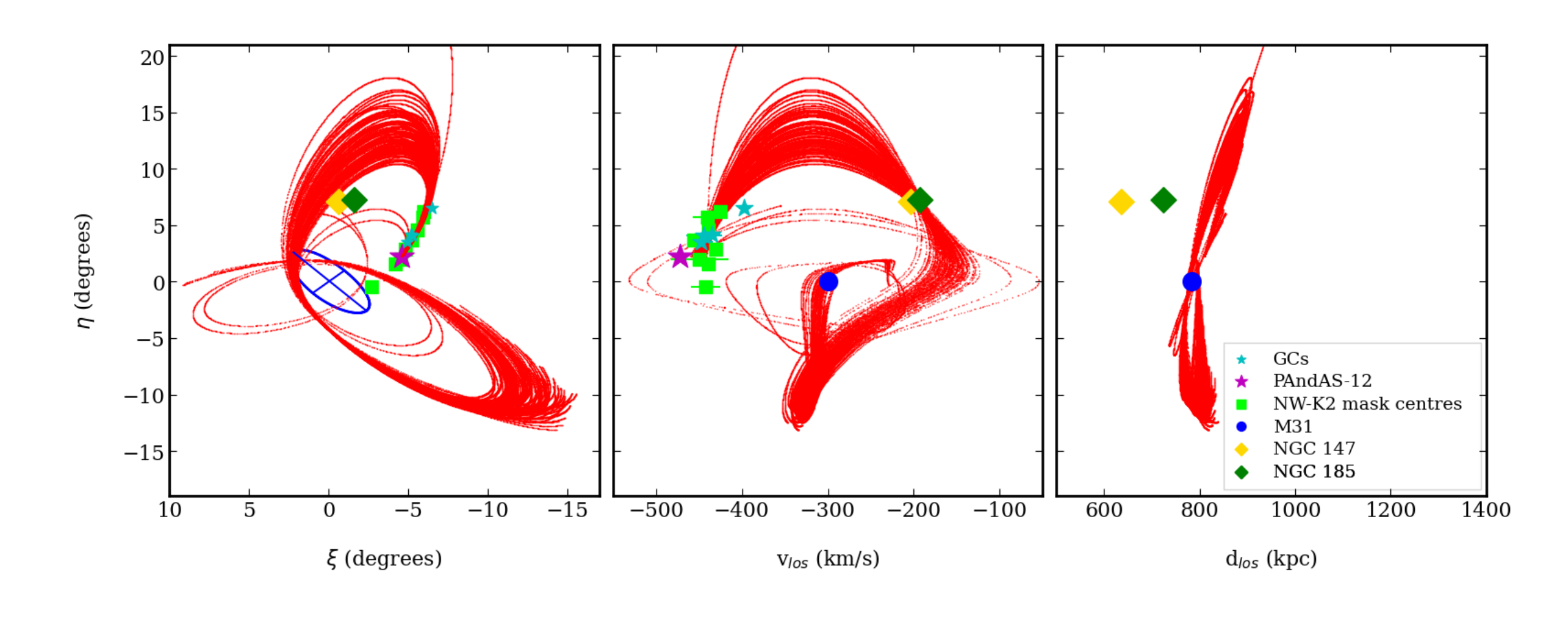}
    	\vspace*{-6mm}\caption[Model orbits for NW-K2 including properties of NGC 147 and NGC 185]
	{Model orbits for NW-K2 generated over a 5 Gyr lookback time with the stream moving in a direction towards M31, with the orbit terminating at PAndAS-12 (the magenta star). In all three plots the green rectangles represent the observed masks along NW-K2, cyan stars represent the properties of the other GCs associated with NW-K2.  The properties of NGC 147 are indicated by the green diamond and those for NGC 185 by a yellow diamond. The left hand plot shows the on-sky position of 100 random orbits for the stream track which, in the early stage of the orbit extends out to the location of NGC 147 and NGC 185.  The middle plot shows the line of sight velocities for the orbits, GCs and dwarf galaxies. The right hand plot indicates the line of sight distances of the stream, the dwarf galaxies and M31 (represented by the blue circle).}
	\label{OM_Fig31}
\end{figure*}

To explore a suggestion in \cite{RefWorks:516} and \cite{RefWorks:449} that NW-K2 could extend round in the general direction of the dwarf elliptical galaxies NGC 147 and NGC 185, we note that our best fit orbits do not obviously suggest such a connection.  So, we plot the back projection of the 100 random orbits to see if there are any configurations of the orbit that could intersect with NGC 147 and/or NGC 185.  These additional results, shown in  Figure \ref{OM_Fig31}, are also not conclusive of a connection between NW-K2 and NGC 147 or NCG  185 aside from similarities in velocities. So while we could hypothesise that there is unlikely to be a connection between NW-K2 and NGC 147/NGC185, to form a more substantive view we would need to compare model orbits for them all to see how they match up over similar time frames.

\section{Conclusions} \label{OM_Conclusions} 
We present the first dynamical stream fits for NW-K1 along with our predictions for the proper motions of streams NW-K1 and NW-K2. Our models, 
while subject to the approximate validity of the adopted fixed potential of M31, produce good representations of the stream tracks for NW-K1 and NW-K2 providing line of sight locations and radial velocities that are good matches to the observational data.  This enables us to conclude that both streams are moving towards M31, speeding up as they move in a south-easterly direction both on the sky and in 3d.  Both appear to lie behind M31 and are furthest away at their north-west extents.  And XXVII remains a plausible candidate for the progenitor of NW-K1 and is most likely in the final throes of tidal disruption, while the progenitor of NW-K2 could be either completely disrupted or lie where the stream disappears into M31's halo.

Reviewing our model orbits for NW-K1 and NW-K2 we find that there are no configurations indicating any connection between the two streams. We, therefore, conclude that they are separate features and that the NW stream is not a single structure. We recommend that NW-K1 be renamed the \enquote{Andromeda XXVII Stream} and that NW-K2, now one of the longest structures around M31 in its own right, retains the name the \enquote{North West Stream}.

Further investigation of these intriguing stellar streams, with their dSph progenitor and co-located GCs, has the potential to provide useful insights into how M31 was formed and how it continues to evolve.  However,  lying as they do at the limits of M31's outer halo, they are almost too far and too faint, for current technology to provide much more detail than we already have.  Still, our simple model orbits could provide useful guides as to where to search for more stream members along the predicted trajectories, as has been done for the MW.  This would enable us to greatly increase our stream datasets and facilitate more detailed modelling of their tracks and, by probing deeper into the luminosity function, search for breaks in the streams.  \textit{N}-body simulations of these streams, in conjunction with those already done for other streams and shelves around M31, could provide us with the multi stream kinematics to help constrain M31's dark matter halo. There is also every possibility that future surveys, such as DESI, which has already started to explore the accretion history of M31, \cite{M:977}, could be extended to map the remainder of \mbox{PAndAS} footprint around M31 and that the Subaru Prime Focus Spectrograph; the Maunakea Spectroscopic Explorer; the James Webb Telescope (JWST); the Nancy Grace Roman Telescope or the Thirty Meter Telescope will enable us, eventually, to unravel the mystery of Andromeda XXVII and its north west streams.

\begin{table*}
	\centering
	\setlength\extrarowheight{2pt}
	\caption[Predicted proper motions for M31, And XXVII, NW-K1 and NW-K2]
	{Predicted proper motions for M31, NW-K1 and NW-K2. The table includes our predications for the proper motions of NW-K1 and NW-K2, derived from the fitting the stream using the mask centres, together with values for M31 from other works using measured data.  \cite{RefWorks:459} used data from the second \textit{Gaia} data release (\textit{Gaia}DR2) catalogue to determine the values for the M31 proper motions.  The proper motions derived by  \cite{M:838} used the Early Third \textit{Gaia} data release catalogue (\textit{Gaia} EDR3).
	}		
	\label{OM_table:4}
	\begin{tabular}{lcc} 
		\hline
	Source & ${\mu^*_{\alpha}}$ (mas/yr) & $\mu$$_{\delta}$ (mas/yr) \\
		\hline 

NW-K1 - towards M31             &  0.078$^{+0.015}_{-0.012}$           &   $-$0.05$^{+0.008}_{-0.009}$   \\[0.75ex]
NW-K2 - towards M31             &  0.085$^{+0.001}_{-0.002}$           &   $-$0.095$^{+0.003}_{-0.004}$  \\[0.75ex]

\\							
M31& & \\
 \cite{RefWorks:459}                           & 0.065   $\pm$ 0.018            &  $-$0.057 $\pm$ 0.015                 \\[0.5ex]
 \cite{M:838}                                       & 0.0489   $\pm$ 0.011           &  $-$0.037 $\pm$ 0.008                \\[0.5ex]
 
		\hline
	\end{tabular}
\end{table*}

\section{Acknowledgements}

The authors wish to thank the anonymous reviewer for their insightful comments and advice. JP wishes to thank Barry Sullivan, Joan Sullivan and Stuart Sullivan for their inspiration.  JP also wishes to thank Mark Fardal, Dougal Mackey and Alan McConnachie for their donations of data and private communications.

This work used the community-developed software packages: Matplotlib (\citealt{RefWorks:616}), NumPy (\citealt{RefWorks:615}) and Astropy (\citealt{RefWorks:613, RefWorks:614, M:1115}) and Uncertainties (\citealt{M:1031}).

Most of the observed data presented herein were obtained at the W.M. Keck Observatory, which is operated as a scientific partnership among the California Institute of Technology, the University of California and the National Aeronautics and Space Administration. The Observatory was made possible by the generous financial support of the W.M. Keck Foundation. Data were also used from observations obtained with MegaPrime/MegaCam, a joint project of CFHT and CEA/DAPNIA, at the Canada-France-Hawaii Telescope which is operated by the National Research Council of Canada, the Institut National des Sciences de l'Univers of the Centre National de la Recherche Scientifique of France, and the University of Hawaii. The authors wish to recognise and acknowledge the very significant cultural role and reverence that the summit of Mauna Kea has always had within the indigenous Hawaiian community. 
\\
\\
\section{Data Availability}
The data used in this paper are available herein, in P19 and in their associated on-line supplementary materials and in \cite{M:1147} and on-line supplementary materials. The raw DEIMOS data are available via the Keck archive.\\

\bibliography{BiblogNWK1}
\bibliographystyle{mnras}

\clearpage
\appendix
\onecolumn
\section{Properties of stream NW-K1 and NW-K2 stars}\label{Properties of stream stars}

These tables show the  properties of the stars associated with streams NW-K1 and NW-K2. The columns include: (1) Star number; (2) Right Ascension in J2000; (3) Declination in J2000; (4) \textit{i}-band magnitude; (5) \textit{g}-band magnitude; (6) line of sight heliocentric velocity, \textit v.
\\
\begin{table*}[h]
	\centering	
	\setlength\extrarowheight{1pt}
	\vspace*{-5mm}
	\caption{Properties of the NW-K1 stream stars from P19.}
	\label{table:A1} 
	\begin{tabular}{lccccc} 
		\hline
		Mask/star  & $\alpha$ & $\delta$ & \textit{i} & \textit{g} &  \multicolumn{1}{c}{\textit{v}}   \\ 
                  &  $hh$ : $mm$ : $ss$     & $^o$ : $^{\prime}$ : $^{\prime \prime}$ &&&  \kms  \\ [0.5ex]
		\hline
7And27 & & & & &     \\ 
 5  & 00:37:18.84 & +45:23:19.3 &  22.1  &  23.3  &   $-$463.2 $\pm$ 4.8   \\
 6  & 00:37:19.75 & +45:23:51.8 &  21.4  &  22.8  &   $-$539.6 $\pm$ 4.0   \\
 7  & 00:37:19.82 & +45:24:17.8 &  21.6  &  23.0  &   $-$476.1 $\pm$ 6.4   \\
11  & 00:37:19.24 & +45:21:36.4 &  21.8  &  23.1  &   $-$563.1 $\pm$ 4.1   \\
19  & 00:37:33.79 & +45:25:18.9 &  22.4  &  23.6  &   $-$539.2 $\pm$ 9.6   \\
20  & 00:37:41.84 & +45:25:27.9 &  22.0  &  23.5  &   $-$544.7 $\pm$ 5.2   \\
31  & 00:37:36.90 & +45:27:06.8  &  21.8  &  23.3  &  $-$533.3 $\pm$ 5.3   \\
32  & 00:37:43.68 & +45:27:11.3 &  21.3  &  23.0  &   $-$533.5 $\pm$ 3.2   \\
46  & 00:37:07.86 & +45:22:50.5 &  22.6  &  23.8  &    $-$579.1 $\pm$ 15.6  \\
54  & 00:37:21.21 & +45:24:25.2 &  23.2  &  24.1  &  $-$499.5 $\pm$ 11.4  \\
55  & 00:37:19.59 & +45:24:37.5 &  23.2  &  24.2  &   $-$480.4 $\pm$ 4.0   \\
\\
A27sf1 & & & & &     \\ 
11  & 00:39:28.22 & +45:09:7.8  &  21.3  &  23.0  &  $-$540.9 $\pm$ 3.2   \\
23  & 00:40:12.21 & +45:04:21.1 &  22.1  &  23.4  &  $-$535.8 $\pm$ 7.6   \\
31  & 00:39:37.65 & +45:8 :52.8 &  22.8  &  24.0  &  $-$560.0 $\pm$ 16.5  \\
32  & 00:39:24.05 & +45:10:26.6 &  22.9  &  23.9  &  $-$571.2 $\pm$ 5.6   \\
33  & 00:39:18.52 & +45:10:29.4 &  22.8  &  24.0  &  $-$525.1 $\pm$ 12.1  \\
35  & 00:39:30.70 & +45:11:07.3  &  22.7  &  24.0  &  $-$517.3 $\pm$ 11.0   \\
37  & 00:39:22.19 & +45:11:56.9 &  22.6  &  24.0  &  $-$540.9 $\pm$ 9.6   \\
38  & 00:39:25.79 & +45:12:53.9 &  22.6  &  23.9  &  $-$549.1 $\pm$ 16.7  \\
\\		
603HaS & & & & &     \\ 
10  & 00:39:08.53 & +45:15:46.8 &  21.2  &  23.0  &  $-$527.9 $\pm$ 5.0  \\
15  & 00:39:05.93 & +45:16:55.3 &  22.1  &  23.4  &  $-$526.2 $\pm$ 3.1   \\
20  & 00:39:25.73 & +45:19:55.0 &  22.4  &  23.6  &  $-$531.5 $\pm$ 4.0    \\
32  & 00:38:30.15 & +45:18:18.1 &  21.7  &  23.1  &  $-$519.6 $\pm$ 6.0   \\
33  & 00:38:46.02 & +45:17:28.4 &  21.2  &  23.2  &  $-$537.5 $\pm$ 4.0    \\
35  & 00:38:38.75 & +45:17:33.4 &  21.7  &  23.2  &  $-$546.9 $\pm$ 8.1   \\
38  & 00:38:44.39 & +45:15:36.2 &  21.7  &  23.3  &  $-$526.2 $\pm$ 5.2   \\
46  & 00:38:32.03 & +45:19:34.1 &  22.3  &  23.7  &  $-$535.8 $\pm$ 17.3  \\
\\
A27sf2 & & & & &     \\ 
33  & 00:36:04.85 & +45:31:17.8 &  22.6  &  24.0  &  $-$521.6 $\pm$ 15.5  \\
42  & 00:36:42.08 & +45:34:11.6 &  22.5  &  23.7  &  $-$517.0 $\pm$ 8.3    \\
 \\
604HaS & & & & &     \\  
16  & 00:31:44.15 & +46:11:9.8  &  21.6  &  23.0  &  $-$503.1 $\pm$ 9.1   \\
 \\ 
A27sf3 & & & & &     \\ 
 7  & 00:30:33.80 & +46:10:30.1 &  21.6      &  23.1  &  $-$489.1 $\pm$ 7.9    \\
31  & 00:30:33.67 & +46:13:04.5  &  22.2  &  23.6  &  $-$505.8 $\pm$ 7.3   \\
32  & 00:30:45.72 & +46:13:25.7 &  22.3  &  23.7  &  $-$500.4 $\pm$ 11.3  \\
35  & 00:30:32.56 & +46:14:45.8 &  22.6  &  23.7  &  $-$498.7 $\pm$ 3.8   \\
		\hline
		\vspace*{-6.0mm}
	\end{tabular}
\end{table*}

\begin{table*}[h]
\vspace*{5mm}
\centering	
\setlength\extrarowheight{1pt}	
\vspace*{-5mm}
\caption{Properties of the NW-K2 stream stars from \protect\cite{M:1147} .
}
\label{table:A2} 
\begin{tabular}{lccccc}
\hline
Mask/star  & $\alpha$ & $\delta$ & \textit{i} & \textit{g} &   \multicolumn{1}{c}{\textit{v} }  \\ 
                  &  $hh$ : $mm$ : $ss$     & $^o$ : $^{\prime}$ : $^{\prime \prime}$ &&&   \kms   \\ [0.5ex]
\hline 
NWS6 & & & & &     \\ 
 2  & 00:28:54.13 & +40:43:32.5 &  22.5  &  23.7  &   $-$472.3 $\pm$ 8.5    \\
 3  & 00:28:39.36 & +40:43:51.1 &  22.3  &  23.5  &  $-$443.1 $\pm$ 6.2     \\
 8  & 00:28:08.56 & +40:44:44.8 &  21.4  &  23.6  &  $-$459.8 $\pm$ 17.6     \\
21  & 00:28:19.34 & +40:45:54.7 &  23.3  &  24.3  & $-$438.2 $\pm$ 8.8     \\
24  & 00:28:37.95 & +40:46:11.6 &  22.9  &  24.0  &  $-$410.4 $\pm$ 5.3      \\
26  & 00:28:43.14 & +40:46:35.3 &  22.7  &  24.0  &  $-$416.0 $\pm$ 6.9      \\
39  & 00:28:04.05   & +40:48:29.8 &  22.4  &  23.6  &  $-$402.7 $\pm$ 11.2       \\
42  & 00:28:31.57 & +40:44:32.9 &  21.9  &  23.1  &  $-$420.3 $\pm$ 6.1      \\
46  & 00:27:50.43 & +40:45:04.9  &  21.3  &  22.8  &  $-$450.0 $\pm$ 3.6      \\
56  & 00:28:41.35 & +40:46:01.6  &  21.5  &  23.5  &  $-$413.9 $\pm$ 6.2      \\
58  & 00:29:01.52 & +40:46:07.2  &  21.4  &  22.9  &  $-$434.3 $\pm$ 4.7      \\
\\
NWS5 & & & &      \\ 
 1   & 00:19:22.93 & +42:41:01.4  &  22.3  &  23.5  &  $-$458.2 $\pm$ 9.0    \\
 7  & 00:19:59.59 & +42:42:54.1 &  23.3  &  24.4  &  $-$421.3 $\pm$ 6.1      \\
 8  & 00:19:42.04 & +42:43:04.9  &  22.8  &  24.0  &  $-$434.7 $\pm$ 11.4     \\
16  & 00:19:48.46 & +42:44:11.3 &  22.8  &  23.9  &  $-$477.0 $\pm$ 11.9     \\
20  & 00:20:40.26 & +42:44:38.1 &  21.2  &  23.0  &  $-$429.5 $\pm$ 2.4     \\
23  & 00:20:38.04 & +42:45:03.2  &  22.0  &  23.3  &  $-$434.5 $\pm$ 6.4      \\
\\
507HaS & & & &      \\ 
17  & 00:18:09.16   &   +43:07:36.5  & 22.4  & 24.0  &    $-$406.7 $\pm$14.4     \\
26  & 00:17:30.26 & +43:08:39.6 &  22.2  &  23.5  &   $-$463.2 $\pm$ 14.9    \\
49  & 00:17:29.70   & +43:04:51.0 &  22.1  &  23.4  &   $-$449.6 $\pm$ 4.0      \\
\\
506HaS & & & & &     \\ 
46  & 00:15:39.35 & +44:00:00.9   &  21.9  &  23.2  &   $-$430.2 $\pm$ 5.0     \\
49  & 00:15:15.58 & +43:57:00.8   &  21.7  &  23.3  &   $-$424.1 $\pm$ 3.7     \\
50  & 00:15:25.92 & +43:58:55.9 &  22.0  &  23.5  &   $-$440.4 $\pm$ 5.3      \\
\\
NWS3 & & & &      \\ 
 6  & 00:13:06.69 & +44:38:10.7 &  21.9  &  23.5  &   $-$446.3 $\pm$ 5.4         \\
14  & 00:13:19.47 & +44:40:26.2 &  22.1  &  23.5  &   $-$461.9 $\pm$ 8.1       \\
23  & 00:13:33.13 & +44:43:02.3  &  22.9  &  24.3  &   $-$445.1 $\pm$ 10.4     \\
24  & 00:13:40.37 & +44:44:08.2  &  21.3  &  23.0  &   $-$461.8 $\pm$ 5.9       \\
28  & 00:13:43.67 & +44:45:27.5 &  21.8  &  23.4  &   $-$429.6 $\pm$ 8.1       \\
29  & 00:13:32.96 & +44:45:41.8 &  22.4  &  23.9  &   $-$466.6 $\pm$ 10.1    \\
33  & 00:13:35.29 & +44:47:28.1 &  23.2  &  24.6  &   $-$456.4 $\pm$ 9.8           \\
43  & 00:13:55.01 & +44:49:41.3 &  22.6  &  23.9  &   $-$465.0 $\pm$ 19.0    \\
\\
704HaS & & & &      \\ 
13  & 00:10:33.23 & +45:31:16.9 &  22.5  &  23.7  &   $-$444.8 $\pm$ 5.2      \\
30  & 00:11:28.49 & +45:32:10.4 &  21.5  &  23.2  &   $-$425.5 $\pm$ 3.0      \\
45  & 00:11:34.06 & +45:33:17.8 &  21.5  &  23.2  &   $-$446.5 $\pm$ 3.5      \\
61  & 00:10:44.95 & +45:34:19.9 &  22.1  &  23.5  &   $-$437.2 $\pm$ 4.3      \\
\\
606HaS & & & &      \\ 
42  & 00:08:14.30 & +46:37:57.9 &  21.2  &  22.8  &   $-$438.4 $\pm$ 5.0      \\
\\
NWS1 & & & & &    \\ 
 6  & 00:7 :31.97 & +47:03:08.5  &  22.6  &  23.9  &   $-$437.1 $\pm$ 4.8     \\
15  & 00:7 :45.43 & +47:06:39.7 &  22.5  &  23.9  &   $-$437.6 $\pm$ 8.9      \\
27  & 00:8 :11.01 & +47:11:50.1 &  22.6  &  23.8  &   $-$418.0 $\pm$ 11.7     \\

\end{tabular}
\end{table*} 


\end{document}